\documentstyle[12pt,cite]{article}
\def\thebibliography#1{\section*{References\@mkboth
  {REFERENCES}{REFERENCES}}\list
  {[\arabic{enumi}]}{\settowidth\labelwidth{[#1]}\leftmargin\labelwidth
    \advance\leftmargin\labelsep
    \usecounter{enumi}}
    \def\newblock{\hskip .11em plus .33em minus -.07em}
    \sloppy
    \sfcode`\.=1000\relax}

\def\figcap{\section*{Figure Captions\markboth
        {FIGURECAPTIONS}{FIGURECAPTIONS}}\list
        {Figure \arabic{enumi}:\hfill}{\settowidth\labelwidth{Figure
99:}
        \leftmargin\labelwidth
        \advance\leftmargin\labelsep\usecounter{enumi}}}
 \relax
\def\tabcap{\section*{Tables\markboth
        {TABLES}{TABLES}}\list
        {Table \arabic{enumi}:\hfill}{\settowidth\labelwidth{Table
9:}
        \leftmargin\labelwidth
        \advance\leftmargin\labelsep\usecounter{enumi}}}
 \relax
\newskip\humongous \humongous=0pt plus 1000pt minus 1000pt

\newif\ifdtup

\def\nw{{\mbox{${n}_{B}$}}}
\newcommand{\s}[1]{\mbox{\small#1}}
\newcommand{\HERA}{\s{HERA}}
\newcommand{\BLNV}{\s{BLNV}}
\newcommand{\HW}{\s{HERWIG}}
\newcommand{\JS}{\s{JETSET}}

\newcommand{\QCD}{\s{QCD}}

\newcommand{\ATLAS}{\s{ATLAS}}

\newcommand{\LHC}{\s{LHC}}

\newcommand{\LOME}{\s{LOME}}
\newcommand{\CERN}{\s{CERN}}

\newcommand{\RAMBO}{\s{RAMBO}}
\newcommand{\MAMBO}{\s{MAMBO}}
\newcommand{\HERBVI}{\s{HERBVI}}

\newcommand{\uu}[1]{\verb!#1!\endgroup}     

\newcommand{\mb}[1]{\ifmmode#1\else\mbox{$#1$}\fi}

\newcommand{\sss}{\scriptscriptstyle\rm}

\newcommand{\lqcd}{\mb{\Lambda_{\sss QCD}}}

\newcommand{\ee}        {\mb{\rm e^+e^-}}

\newcommand{\bbar}      {\mb{\rm \bar b}}
\newcommand{\tbar}      {\mb{\rm \bar t}}

\newcommand{\Z}         {\mb{\rm Z^0}}
\newcommand{\W}         {\mb{\rm W^\pm}}
\newcommand{\Wp}        {\mb{\rm W^+}}
\newcommand{\Wm}        {\mb{\rm W^-}}

\newcommand\as{\mb{\alpha_s}}
\newcommand\be{\mb{\beta}}

\newcommand\up{\mb{\upsilon}}

\newcommand\calC{\mb{{\cal C}}}

\newcommand{\gtap}{\;\raisebox{-.4ex}{\rlap{$\sim$}} \raisebox{.4ex}{$>$}\;}

\newcommand{\MC}{Monte Carlo}

\newcommand{\VEV}[1]{\mb{\left\langle #1\right\rangle}}

\newcommand{\barb}{\mb{\bar\beta}}
\font\bigm=cmex10 scaled\magstep1
\def\bigint_#1^#2{\hbox{\raise16.8pt\hbox{\bigm\char'132}}_{#1}^{\;\;#2}}

\font\Bigm=cmex10 scaled\magstep2
\def\Bigint_#1^#2{\hbox{\raise19.4pt\hbox{\Bigm\char'132}}_{\!#1}^{\:\;\;#2}}

\font\biggm=cmex10 scaled\magstep3
\def\biggint_#1^#2{\hbox{\raise23.2pt\hbox{\biggm\char'132}}_{\!#1}^{\;\;\;#2}}

\font\Biggm=cmex10 scaled\magstep4
\def\Biggint_#1^#2{\hbox{\raise28.4pt\hbox{\Biggm\char'132}}_{\!\!#1}^{\:\;\;\;#2}}

\font\BIGm=cmex10 scaled\magstep5
\def\BIGint_#1^#2{\hbox{\raise33.5pt\hbox{\BIGm\char'132}}_{\!\!\!#1}^{\:\;\;\;\;#2}}

\font\normalm=cmex10
\def\normalint_#1^#2{\hbox{\raise15pt\hbox{\normalm\char'132}}_{#1}^{\;\;#2}}

\def\smallint_#1^#2{\hbox{\raise10pt\hbox{\bigm\char'122}}_{#1}^{\;#2}}

\newdimen\mycirclesize
\newcount\mycirclesiz

\def\mycircl#1{%
\mycirclesiz\mycirclesize%
\mycirc{\the\mycirclesiz}%
}
\def\mycirc#1{%
\special{" newpath 0 0 #1 65536 div 0 360 arc stroke }%
}
\newcommand\su[1]{\mb{{\mathrm{SU}}(#1)}}

\def\enplu{\mb{E_{l^+}}}
\def\enmin{\mb{E_{l^-}}}

\begin{document}
\bibliographystyle{bibstyle}
\begin{titlepage}
\begin{flushright}
     CERN--TH.7090/93 \\
     Cavendish--HEP--93/6
     \end{flushright}
\vspace*{\fill}
\begin{center}
{\Large \bf Monte Carlo Simulation of Baryon and Lepton \\
Number Violating Processes at High Energies\footnote{
Research supported in part by the UK Science and Engineering Research
Council and by the EC Programme ``Human Capital and Mobility", Network
``Physics at High Energy Colliders", contract CHRX-CT93-0537 (DG 12 COMA).}
}
\end{center}
\par \vskip 2mm
\begin{center}
        {\bf M.J.\ Gibbs} \\
        Cavendish Laboratory, University of Cambridge \\
        Madingley Road, Cambridge CB3 0HE, U.K.
        \par \vskip 2mm \noindent
        {\bf A.\ Ringwald,
        \footnote{Present address: DESY, Hamburg, Germany.}
        B.R.\ Webber
        \footnote{Present address: Cavendish Laboratory, Cambridge, UK.}} \\
        Theory Division, CERN, CH-1211 Geneva 23, Switzerland
        \par\noindent 
        and
        \par\noindent
        {\bf J.T.\ Zadrozny
        \footnote{Present address: 31 Moran Avenue,
        Princeton, New Jersey, USA.}}\\
        Cavendish Laboratory, University of Cambridge \\
        Madingley Road, Cambridge CB3 0HE, U.K.
\end{center}
\par \vskip 2mm

\begin{center} {\large \bf Abstract} \end{center}
\begin{quote}
We report results obtained with the first complete event generator
for electroweak
baryon and lepton number violating interactions at supercolliders.
Typical events contain of the order of 50 electroweak gauge bosons,
some Higgs bosons and quarks and leptons of all generations. There is
still great uncertainty about the expected rate, but an event
generator is needed in any case to establish what experimental limits
can be placed on the cross section, and to determine whether, even
if such spectacular events are seen, baryon and/or lepton number
violation can be conclusively demonstrated. We find that baryon
number violation would be very difficult to establish, but lepton
number violation can be seen provided at least 
a few hundred L violating
events are available with good 
electron or muon identification in the energy
range 10 GeV to 1 TeV. The event generator, which takes the form
of a package (\HERBVI) interfacing to the existing simulation program
\HW, should be useful for the coming period of
detailed experiment design for the Large
Hadron Collider (\LHC) at \CERN.
\end{quote}
\vspace*{\fill}
\begin{flushleft}
     CERN--TH.7090/93 \\
     \today
\end{flushleft}
\end{titlepage}

\newpage
\thispagestyle{empty}
\vspace*{10.0cm}
\setcounter{page}{0}%

\newpage
 
\section{Introduction}
 
Current experiments have given
us no evidence for physics beyond the Standard
Model. Indeed, the Standard Model may be valid, as an effective
theory, up to very high energies, say the Planck mass, $10^{19}$ 
GeV, if the Higgs mass is below several hundreds of GeV. 
Even this case, 
however, does not necessarily imply that no new phenomena will be
seen in the multi--TeV range which will be explored by future
colliders such as the \CERN\ \LHC. 
There is the intriguing possibility that, above a parton--parton
centre of mass threshold in the multi--TeV range, the cross section
for the nonperturbative production of many, ${\cal O}
(\alpha_W^{-1})$, weak bosons W(Z) may be observably large. 
Unfortunately, there is only circumstantial evidence for this to happen
which is, to a large extent, just based on the observation that 
leading order
perturbative calculations for the
production 
of ${\cal O}(\alpha_W^{-1})$ weak bosons 
violate unitarity near the threshold of 
${\cal O}(\alpha_W^{-1} m_W )$~TeV.
This happens both for processes with~
\cite{ri90,es90,mc90} and without~\cite{co90,gp90,gp92}
baryon and lepton number violation (\BLNV).
At present
it is an open question whether the {\it actual}
(beyond perturbation theory) multi-W(Z) cross
sections
become observably large at such multiplicities and energies
(for recent reviews see~\cite{ma92,tin92,ri93}). New theoretical methods
are needed to answer this important question.
 
In the meantime one can contemplate the prospects of
settling the issue experimentally.
Multi-W(Z) processes at the \LHC\ would be clearly distinguishable
from any other Standard Model process~\cite{ri91b}.
This is due to the
hadronic and leptonic decays of the W's and Z's, which lead to
hundreds of charged hadrons and photons with transverse momenta
in the GeV range, and to tens of prompt leptons with transverse
momenta in the tens of GeV range. However, the question whether
\BLNV\ can be demonstrated in such a multi-particle
environment
has not been answered conclusively~\cite{ri91b,fa90}.
It is the purpose of this paper to investigate this question.
 
Using the results of 
leading--order 
\BLNV\ calculations,
it is possible to create a phenomenological model of these
processes, with variable parameters to represent the main
sources of theoretical uncertainty.
In this paper such a model is postulated, and
implemented using a \MC\ event generator.
The program, which takes the form of a 
package \HERBVI~\cite{gi94a} operating
within the environment of the \HW\ event generator,
also allows the generation of B and L 
conserving multi--W(Z) events.
With this package computer
simulations can be performed which permit the feasibility of
demonstrating \BLNV\ to be investigated.
 
The layout of this paper is as follows.
In the remainder of this section we describe
the theoretical basis of electroweak \BLNV.
Section \ref{sec:model} is a discussion of our
approach to modelling \BLNV\ processes at supercolliders,
based on the (rather uncertain) theoretical expectations,
and of practical issues that arise
in the construction of an event generator.  
We formulate two contrasting \BLNV\ models, one based
on leading-order matrix elements (\LOME) and one using
a simpler phase-space approach with fixed boson multiplicity.
We also discuss the possible backgrounds
to \BLNV, and how we model the dominant
background, which is expected to consist of
B and L  conserving multi--W(Z) events.

Section \ref{sec:MCres} contains the results of our
\MC\ studies. We present results for the
simpler model at three energies, including that
proposed for the \LHC.  The \LOME\ model requires
a higher energy and so in that case we show
results only at 40 TeV.
We present our conclusions in Sect.~\ref{sec:conc}.
Brief details of the \HERBVI\ package,
used to obtain the majority of the results
in Sect.~\ref{sec:MCres}, are contained in an appendix.
 
\subsection{Instanton-induced B and L violation}
\label{sec:insint}

Owing to the chiral anomaly~\cite{ad69,bj69,ba69} 
and the V--A structure of weak interactions,
B and L are not strictly conserved in the Standard 
Model~\cite{th76a,th76b}. In 
the presence of nontrivial SU(2) gauge fields 
$W_i$, the fermionic quantum numbers change according to
\begin{equation}
\label{eq:sel}
\triangle L_e =\triangle L_\mu =\triangle L_\tau =
{1\over 3}\triangle B = 
-\triangle 
N_{\rm CS} ,
\end{equation}
where
\begin{equation}
\label{eq:csn}
N_{\rm 
CS} 
\equiv {\alpha_W\over 4\pi}\int d^3x\ \epsilon^{ijk}\ {\rm tr}
\Biggl( F_{ij}W_k - {2ig\over 3} W_iW_jW_k\Biggr) 
\end{equation}
denotes the Chern--Simons number of the gauge field, $F_{ij}$ is the
SU(2) field strength, and $\alpha_W\equiv g^2/(4\pi )$ the weak
fine structure constant. As is suggested by eqs. (\ref{eq:sel})
and (\ref{eq:csn}), strong, nonperturbative gauge fields, of
${\cal O}( 
g^{-1})$, are needed in order to change the Chern--Simons number, or,
equivalently, the fermion numbers, by an integer amount. This is
reflected by the fact that there exists, on topological grounds,  
an energy barrier~\cite{jr76,cdg76}
between gauge fields whose Chern--Simons numbers differ by an integer.
The minimum barrier height is given by the energy of a static saddle--point
solution, the so called ``sphaleron"~\cite{ma83,km84}, which depends
slightly on the Higgs mass, 
\begin{equation}
\label{eq:spm}
M_{\rm sp} = 
2\ B(m_H/m_W)\  {m_W\over \alpha_W}\ \simeq\;\; 	
\mbox{7--14 TeV} \;,
\end{equation}
where the 
parameter $B$ is restricted to lie in the range
$1.57\leq B\leq 2.72$. 
At low energies ($\ll M_{\rm sp}$), therefore,  
anomalous \BLNV\ processes
are only possible by quantum tunnelling under the topological barrier, 
i.e. the corresponding
amplitudes are exponentially suppressed by a tunnelling factor, 
\begin{equation}
\label{eq:tunfac}
{\cal A}_{E\ll M_{\rm sp}}^{\rm BV} \propto
{\rm e}^{-2\pi /\alpha_W} \sim 10^{-78} ,
\end{equation}
which leads to unobservably small cross sections or decay 
rates~\cite{th76a,th76b}.

The amplitude (\ref{eq:tunfac}) is expected to be enhanced when
the process involves a large number of gauge and/or Higgs bosons.
Consider the following \BLNV\ process, which might be observable
in high-energy proton-proton collisions:
\begin{equation}
\label{eq:anoproc}
q~+~q\rightarrow 7~\bar q~+~3~\bar l
~+~n_B~\mbox{W(Z)}~+~n_H~\mbox{H} .
\end{equation}
The amplitude for this process can be estimated by means of the
instanton approach~\cite{th76a,th76b}.    
The instanton is
a time--dependent tunnelling 
solution which interpolates between gauge fields whose Chern--Simons
numbers differ by one unit 
and passes through sphaleron--like fields
for particular values of its collective coordinates.  
It was found  in refs.~\cite{ri90,es90}, 
by expanding the path integral about the instanton, 
that, in 
leading--order ($LO$) in the coupling $\alpha_W$,  
the 
amplitudes for the processes (\ref{eq:anoproc}) grow with 
multiplicity and parton--parton centre of mass energy $\sqrt{\hat s}$  
like      
\begin{equation}
\label{eq:loamp}
{\cal A}_{n_B,n_H\ LO}^{\rm BV} \sim (n_B+n_H)!
~\alpha_W^{(n_B+n_H)/2}
~\mbox{e}^{-2\pi /\alpha_W}
~\Biggl( {\sqrt{\hat s}\over n_B\ m_W}\Biggr)^{n_B}
~m_W^{-(n_B+n_H)} .
\end{equation}
At high energies, these pointlike S-wave amplitudes violate  
unitarity. Owing 
to the factorial growth of the amplitudes~(\ref{eq:loamp})
with the number
of produced bosons, this
violation of unitarity sets in at
multiplicities of ${\cal O}(\alpha_W^{-1})$ and at
parton-parton centre of mass energies of
${\cal O}(\alpha_W^{-1}m_W)$, i.e. the sphaleron energy (\ref{eq:spm}).
 
\begin{figure}[tb]
  \vspace*{8.2cm}
  \includegraphics{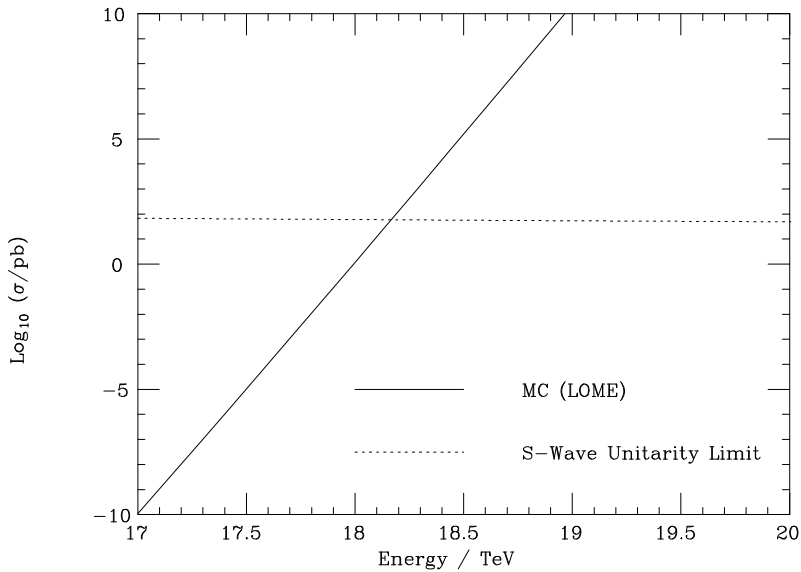}
  \caption[]{Quark-level cross section for 
             leading order \BLNV\ process, solid
         line. The S-wave unitarity limit is shown as a dotted line.}
  \label{fig:qq_lo_res}
\end{figure}

The corresponding
parton level cross section, $\hat{\sigma}_{LO}$, due to
the leading order matrix element (\LOME) of instanton-induced
{\BLNV}~(\ref{eq:anoproc}), is given by~\cite{ri90,es90}
\begin{eqnarray}
{{\hat{\sigma}}_{n_B,n_H\ LO}^{\rm BV}
} & = & \tilde{C}{\cal G}^2 2^n
\upsilon^{-2n} \left[ \frac{
\Gamma \left( n+ \frac{103}{12} \right)}
{\Gamma \left( \frac{103}{12} \right)} \right]^2
\frac{1}{n_B! n_H!} \nonumber \\
& &\times \int \prod_{i=1}^{10}
\frac {{\mathrm d^3}p_i}{\left( 2\pi \right)^3 2
E_i} E_i
\prod_{j=1}^{n_B} \frac {{\mathrm d^3}p_j}{\left( 2\pi \right)^3 2
E_j} \frac{2\left( 4E_j^2 - m_W^2 \right)}{m_W^2}
\nonumber \\
& &\times \prod_{k=1}^{n_H}
\frac{{\mathrm d^3}p_k}{\left( 2\pi \right)^3 2 E_k}
\left( 2\pi \right)^4 \delta^{(4)}
 \left( P_{\mathrm in} - \sum_{i=1}^{10}
 p_i - \sum_{j=1}^{n_B} p_j -\sum_{k=1}^{n_H}
 p_k\right) \;. \label{eq:bvics}
\end{eqnarray}
In this expression $n_B$ is the number
of vector bosons and $n=n_B+n_H$.
The effective coupling
constant ${\cal G}$ is
\begin{equation}
{\cal{G}} = 1.6 \times 10^{-101}\; {\mathrm{GeV}}^{-14},
\end{equation}
$\up=2 m_W/g$ is the Higgs field vacuum expectation value,
and $\tilde{C}$ is a numerical factor
representing the effect of averaging and summing over incoming
and outgoing states of the fermions.
The leading order result
${\hat{\sigma}_{n_B,n_H\ LO}^{\rm BV}
}$, taken at the dominant multiplicities,  
is plotted in Fig.~\ref{fig:qq_lo_res}. 
As anticipated, it rises rapidly with centre of mass
energy, and violates
the S-wave unitarity bound (shown as a dashed line on the same plot)
\begin{equation}\label{eq:unitbound}
{\hat{\sigma}_{unit}}\left( \hat{s}\right) =\frac{16\pi}{\hat{s}}
\end{equation}
at an energy of order the sphaleron energy. 

A violation of unitarity is, of course, unacceptable and indicates
the importance of higher order corrections. 
There are strong arguments that the higher order
corrections to the
fixed multiplicity amplitudes exponentiate in the total cross section
of \BLNV, such that, to exponential accuracy, the
latter can be written as~\cite{ya90,ar90,krt91,mu91a,mu91b,vo91}:
\begin{equation}
\label{eq:stot}
{\hat \sigma}_{\rm tot}^{\rm BV}
\equiv \sum_{n_B,n_H} {\hat \sigma}^{\rm BV}_{n_B,n_H} 
\propto
 \exp\Biggl\lbrack
{4\pi  \over \alpha_W} F
\biggl( {\sqrt{\hat s}\over M_0 } \biggr)
\Biggr\rbrack
 ,
\end{equation}
where
\begin{equation}
\label{eq:m0}
M_0\equiv \sqrt{6}\pi\frac{m_W}{\alpha_W}
\end{equation}
is of the order of the sphaleron mass (\ref{eq:spm}).
The exponent itself is known only in a low energy expansion whose first
few terms are given by  
~\cite{krt91,za90,po90,kr91,dpe91,mu91c,ar91,dpo91,bs92,sil92,za92a}
\begin{equation}
\label{eq:holygrail}
F
( \epsilon )
=
-1 + {9\over 8}
\epsilon^{4/3} -{9\over 16}
\epsilon^2 +
{\cal{O}}\left( \epsilon^{8/3}\right)
,
\end{equation}
where $\epsilon \equiv \sqrt{\hat s}/M_0$.
From this result the following conclusions can be drawn:
(i) The total cross section for \BLNV, though
small, is growing
exponentially
at $(m_W\ll ) \sqrt{\hat s}\ll M_0$; 
in this energy region, the
cross section is insensitive to the Higgs mass.
(ii) The different
terms in the perturbative expansion of the
`holy grail function' $F$ become comparable in size, and the perturbative
expansion breaks down,
at $\sqrt{\hat s}
\sim M_0$. 

This breakdown of perturbation theory is not surprising.
\BLNV\ processes involve the electroweak gauge fields
penetrating the topological barrier between different vacuum sectors.
As the energy approaches $M_0$, the height of this barrier, the
possibility of crossing the barrier classically arises, and the
expansion based upon the instanton tunnelling solution ceases to be valid.
Unfortunately, nothing is known about the behaviour of the holy grail
function around or above the sphaleron scale. Unitarity arguments
~\cite{za91,ms91,ve92} suggest that its increase
will stop at values of order $F\simeq -0.5$, leading
to unobservably small cross sections for \BLNV. However, 
this question is not settled finally.  
 
To end this section, 
we note that
at low energies the total
cross section~(\ref{eq:stot})
is dominated by multi-W(Z) production
($n_B\sim \alpha_W^{-1}$)
rather
than by multi-Higgs production.
In particular one finds at low energies
\begin{equation}
\label{eq:mul}
\bar n_B\sim {4\pi\over 3
\alpha_W} \left( {9\over 8} \epsilon^{4/3} + {\cal{O}}\left(
\epsilon^{2}\right) \right) ,
\end{equation}
whereas
\begin{equation}
\label{eq:mulH}
\bar n_H\sim {4\pi\over\alpha_W}{3\over 32} \epsilon^2 \; .
\end{equation}

\subsection{Sphaleron decay}
\label{sec:sphal}
%
%
%
%
%
It is interesting to compare the above results for the
instanton-induced process (\ref{eq:anoproc}) with those
from sphaleron decay. Since the sphaleron lies
atop the energy barrier through which tunnelling occurs,
it is expected that
the lowest energy classical field trajectories which violate B and L
will pass (in some sense) close to the sphaleron configuration.
Thus, we expect the behaviour of the sphaleron to give us some indication
of what should be observed in an actual \BLNV\ event.
In particular, the decay products of the sphaleron (which is by nature
unstable) should be similar to the gauge and Higgs particle byproducts
of a \BLNV\ event. Therefore, one again expects~\cite{ag87,amc88} 
that the
dominant \BLNV\ processes will involve ${\cal O} (\alpha_W^{-1})$
weak bosons, simply because sphaleron--like intermediate states will
typically decay into many W's and Z's.

The decay of the Klinkhamer--Manton 
sphaleron has been studied numerically in refs.~\cite{he92,za92}.
The sphaleron field configuration was first discretised, and then
allowed to evolve under the classical Euler--Lagrange field equations
with an imposed condition of spherical symmetry
\cite{ra88,wi77,tu91}.
The restriction to spherical symmetry is permissible because the sole
decay channel of the Klinkhamer--Manton sphaleron is spherically symmetric
\cite{ak89};
it should be noted, however,  that this approach may be used only in the limit
of vanishing weak mixing angle $\theta_{\rm W}$.
As a consequence of the imposed spherical symmetry, it is expected that
equal numbers of $W^+,\,W^-\,\mathrm{and }\, Z^0$ of
any given polarisation
state will be produced after the decay.
This is due to the fact that the spherically symmetric ansatz mixes
spacetime and group indices, so that particle identities may be interchanged
by appropriate spatial rotations.
 
As a sphaleron decays and the energy carried by the gauge fields moves
away from the sphaleron's initial position, the
amplitudes of the gauge fields show the $1/r$ attenuation characteristic
of spherical radiation.
Some time after the sphaleron decay, the amplitudes of the gauge fields
become small enough that the nonlinear terms in the gauge field equations
of motion are negligible.
Thus, a long time after the decay, the various components of the gauge fields
behave approximately like free massive fields, and they evolve independently
of one another.
The multiplicities can then be measured by interpreting the classical
fields in terms of coherent states \cite{kl68,zh90,uzi}.
 
\begin{figure}[tb]
  \vspace*{8.2cm}
  \includegraphics{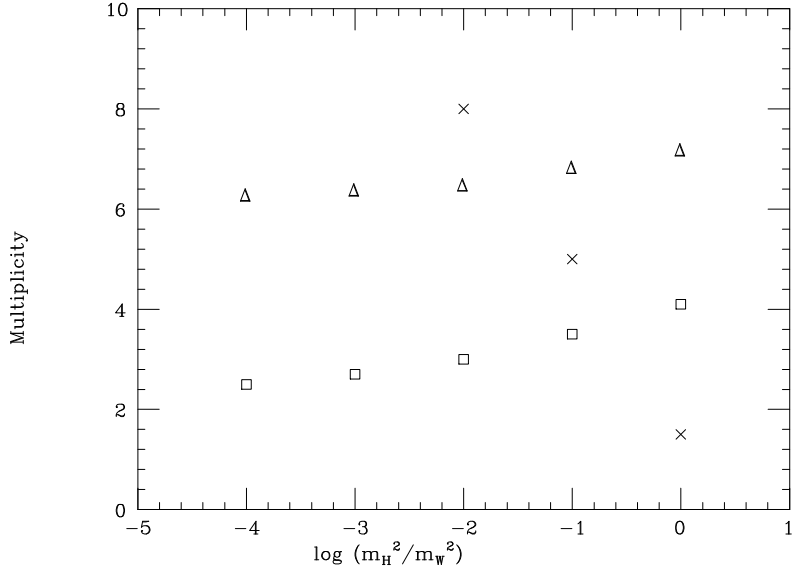}
  \caption[]{Sphaleron decay products 
as a function of the Higgs mass.
Points shown on this graph show the numbers of each kind of particle
in each possible polarisation state:
triangles represent transversely polarised gauge bosons,
squares show longitudinally polarised bosons, and crosses denote
Higgs particles.}
\label{fig:Sphal_dec}
\end{figure}

Figure~\ref{fig:Sphal_dec} shows the expected
particle multiplicities in sphaleron decays for several values
of $m_H^2/m_W^2$.
The triangles indicate the expected numbers of transversely polarised weak
bosons per helicity state and per particle type ($\bar n_T$),
and the squares show the expected numbers of longitudinally polarised
bosons per particle type ($\bar n_L$).
The crosses show the expected numbers of Higgs bosons ($\bar n_H$).
The total number of gauge bosons is then
\begin{equation}
\bar n_B = 6\bar n_T + 3\bar n_L\; .
\label{jtz_totalwmult}
\end{equation}
The expected number of gauge bosons is seen to vary weakly with the
Higgs mass and is of ${\cal O}(\alpha_W^{-1})$, as anticipated. 

\begin{figure}[tb]
  \vspace*{8.2cm}
  \includegraphics{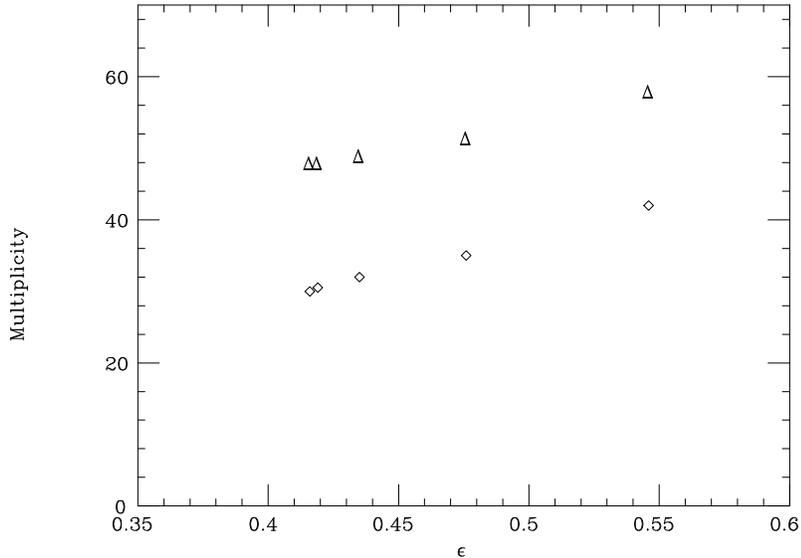}
  \caption[]{Mean number of gauge bosons, $\bar{n}_B$, expected 
to be produced in a baryon
number--violating event, as a function 
of $\epsilon = \sqrt{\hat{s}}/M_0$, where $M_0 = \sqrt{6}\pi m_w
/\alpha_W$. Triangles 
show $\bar{n}^{}_{\rm B}$ from sphaleron decays, and 
diamonds show $\bar{n}^{}_{\rm B}$ from
Eq.~(\ref{eq:mul}).} \label{fig:boson_mult}
\end{figure}

Setting $\sqrt{\hat s} = M_{\rm sp}$ leads to
\begin{equation}
\epsilon_{\rm sp} 
\equiv \frac{M_{\rm sp}}{M_0} = \frac{2}{\sqrt{6} \pi}
     B(m_H/m_W)
\simeq 0.4\ -\ 0.7\;,
\label{jtz_epsilondef}
\end{equation}
which  may be computed readily from the
table of $M_{\rm sp}$ versus $m_H$ provided in \cite{km84}.
Figure~\ref{fig:boson_mult} compares the expected gauge boson multiplicities
computed from (\ref{jtz_totalwmult}), shown by triangles, with
those given by Eq.~(\ref{eq:mul}),
denoted by diamonds.
It can be seen that the multiplicities predicted by
(\ref{eq:mul})
are roughly 30\% smaller than those expected from sphaleron decays.
This discrepancy is not very surprising in view of the fact that the
perturbative expansion of $F(\epsilon)$ does not converge rapidly
for values of $\epsilon$ so near to unity.

\section{Modelling B and L violation}\label{sec:model}

In the absence of a more complete theory at energies of order the  
sphaleron energy, we will
use the gross features of instanton and sphaleron calculations,  
such as the threshold--like behaviour of the cross section and the 
parametrically large vector boson multiplicity, 
as a guide for phenomenological modelling 
of electroweak \BLNV.
The steeply rising cross section means that \BLNV\ processes
have a threshold nature. 
Three parameters are used which encompass
the main features of the processes: a threshold energy $\sqrt{\hat{s}_0}$,
a threshold cross section $\hat{\sigma}_T$, and the boson multiplicity, \nw.
The parton level cross section is then modelled as
a step function
\begin{equation}\label{eq:theta_cs}
\hat{\sigma}\left(\sqrt{\hat{s}}\right) = {\hat{\sigma}_{T}}\;
\theta\left(\sqrt{\hat{s}} - \sqrt{\hat{s}_{0}}\right)
\end{equation}
following the approach used in~\cite{ri91b}. We assume that
in the rest frame of the interacting partons the events are isotropic, and
that the outgoing particles have an energy 
distribution that fills the available phase space
uniformly. We first discuss the model parameters and
general assumptions, and then, in separate subsections, explain the
detailed modelling of events and background.

The magnitude of ${\hat{\sigma}_{T}}$\ is
obviously a crucial
quantity, as it determines the rate at which the interactions
will occur. It is also difficult to estimate because of
the theoretical uncertainties, discussed above.
For this reason we have performed
estimates in terms of the number of events required to verify
the existence of \BLNV, assuming a comparable background of B and L
conserving `multi--W' events with a similar boson multiplicity.
The rationale behind this model for the background is presented
in Sect.~\ref{sec:blback}.

The high energy processes under consideration mean that,
for any $pp$ collider proposed for the near future,
the threshold energy $\sqrt{\hat{s}_0}$ is a significant fraction
of the total beam energies. Consequently, events will be at large
momentum fractions $x$, and also will tend to be 
central within the detector. The exact size of $\sqrt{\hat{s}_0}$ is
not significant for our studies. We have set this parameter at
the point where the instanton calculation violates the S-wave 
unitarity bound, $\sim M_{0}$, except for the studies performed
at 17~TeV where the value of 5~TeV $\left(\sim
2 m_W /\alpha_W\right)$ is used.

\begin{figure}[tb]
  \vspace*{8.2cm}
  \includegraphics{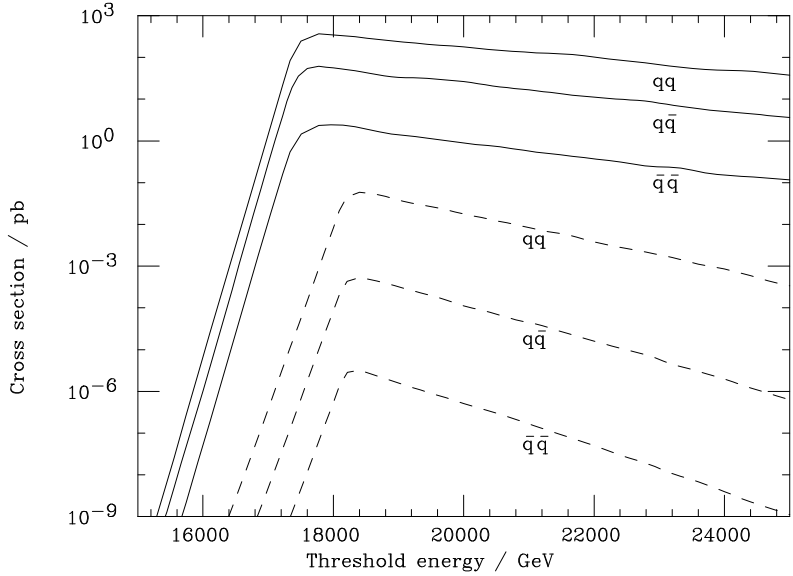}
  \caption[]{Convolution of step threshold function with proton structure
functions for $pp$ energies of 40 TeV (dashed line) and 
200~TeV (solid line). The parton level cross section, $\hat{\sigma}_T$,
has been taken as the minimum of either the value from the
\LOME\ calculation or the S-wave unitarity limit.}
\label{fig:qqconv}
\end{figure}

Convolution of~(\ref{eq:theta_cs})
with the appropriate proton structure functions
allows a comparison of the relative rates of processes
with similar parton-level cross sections mediated by the possible
initial state combinations of $qq$, $q\bar{q}$, and $\bar{q}\bar{q}$ at
a $pp$ collider. These relative rates are plotted versus threshold
energy in Fig.~{\ref{fig:qqconv}}.
The results shown were obtained using the {EHLQ} structure
functions for the proton~\cite{ehkq86}, but the dominance of $qq$
processes at high threshold energies is not sensitive to the
choice of structure functions. It follows simply from the fact
that the valence $u$ and $d$ quark distributions dominate over
the sea quarks and gluons at high values of the parton momentum
fraction $x$.

For \BLNV, which can only occur for incoming $qq$ or $\bar{q}\bar{q}$,
the latter contribution is two to three orders of magnitude smaller than
the former. Therefore the contribution of \BLNV\ with the opposite sign 
(creation of baryons as opposed to antibaryons) is negligible.

In this paper, two alternative methods for generating the
boson multiplicity \nw\ are used. One approach is to
set $n_B = 30 \sim 1/\alpha_W$\, as this is the multiplicity
scale at which the violation of unitarity arises \cite{ri91b}. The
second approach is to distribute \nw\ according to
the \LOME\ expression (\ref{eq:bvics}).\ In either case, the
high multiplicity of electroweak bosons
means that \MC\ techniques that can generate energy and momentum
configurations of many massive particles efficiently have to be 
used. This is discussed in Sect.~\ref{sec:bosfin}.

We now give a justification of the assumption that particles can be
generated isotropically. We saw above that valence quark interactions
are dominant.  The valence quark structure functions at large $x$
have the approximate form
\begin{equation}
f\left( x\right) = \left( 1 - x \right)^{n},\ \ \ 
n \simeq 3 
\end{equation}
and the convolution of these structure functions
with (\ref{eq:theta_cs}) produces events with
$\hat s = x_1 x_2 s$ close to the threshold,
where higher partial waves are
suppressed and so the angular distribution is
approximately isotropic.

We can characterize the available energy per particle
by a proportionality constant \cite{ri91b} 
\begin{equation}
f_{W} = \frac{\sqrt{\hat{s}}}{M_{tot}},
\end{equation}
where 
\begin{equation}
M_{tot} = \sum_{i} m_{i}
\end{equation}
is the total mass of the outgoing particles.
The average kinetic energy per particle can then be estimated 
using 
\begin{equation}\label{eq:avenest}
{\bar{E}} \sim \left( f_{W}-1\right) {\bar{m}}
\end{equation}
for an average particle mass $\bar{m}$.
For values of $f_W$ close to one,
the average kinetic energy per particle is small,
and one can expect that events will be fairly isotropic.
This isotropic character will be reduced for large values of
$f_W$. 

\subsection{Boson final states}\label{sec:bosfin}

Evaluation of the \LOME\ cross section 
(\ref{eq:bvics}) has been performed
using a \MC\ technique.
This approach is closely
related to the generation of events
for \MC\ simulations. Events are generated uniformly in phase
space for a fixed particle 
configuration $\left( n_B, n_H, {\cal C}_f\right)$,
where the symbol $\calC_{f}$\ represents
the fermion configuration.
For each event, a weight that is the product of the matrix element
for the event and the phase space integral for the configuration is
calculated. The average of these weights is then the \MC\ estimate
of the cross section.

The phase space integral is calculated using
the saddle point method \cite{Bykling}.
This method is based upon Laplace 
transforming the integral in order
to remove the delta function, leaving an inverse transform, which
can then be evaluated using a saddle point approximation.
This technique can also be
used to evaluate 
the cross section~(\ref{eq:bvics}); whilst this allows 
computation of the cross section it is
restrictive from an event generation point of view.

For a given energy and particle configuration, the generation of
events is performed using the \MAMBO\ \MC\ algorithm
\cite{mambopre}. This algorithm produces configurations
in phase space obeying energy and momentum constraints. The
algorithm is more efficient than others, such as 
\RAMBO\ \cite{rambo} from which it was derived, because
various parameters are tuned
for each different configuration of particles
and total energy.
The events generated by \MAMBO\ each have an associated weight
$w$. These weights are distributed
in the range 0 to 1, and so the correct unweighted
distribution is obtained by rejecting 
events with $w$ less than a random number
generated in this range.

The evaluation of the matrix element for each event
is the third step in the cross section calculation. 
Computation of the factor 
in~(\ref{eq:bvics}) for each gauge boson,
averaging over the set of generated events, and multiplying
by the phase space integral for the particle configuration, gives
us an estimate of the leading-order cross section 
${\hat{\sigma}}_{LO}\left( n_B,n_H,{\calC}_f\right)$
for a given
set of final state particles. 
The total quark level cross section
is then obtained by
summing over all possible particle configurations.
The particle numbers are constrained by the requirement 
that the total mass of all the outgoing 
particles has to be less
than the energy of the interaction.

In summing over the 
number of vector bosons, the relative number
of {\Z}\ bosons and photons
has to be considered~\cite{gi94b}. The instanton is a field 
configuration in unbroken \su{2} gauge theory; in the
standard model this symmetry is broken to give the physical
particles \W, \Z\ and $\gamma$. Whilst equal
numbers of \Wp, \Wm\ and $W^0$ bosons of the unbroken symmetry are
produced by the instanton interaction, the
$W^0$ bosons 
are projected
onto the \Z\ and $\gamma$ physical states, with
the relative probabilities of 
$\cos^2 \theta_W$ and $\sin^2 \theta_W$ respectively.
The 
probability distribution for $n_{\gamma}$\ photons
and $n_{Z}$\ \Z\ bosons is then described by
a binomial distribution.

\begin{figure}[tb]
  \vspace*{8.2cm}
  \includegraphics{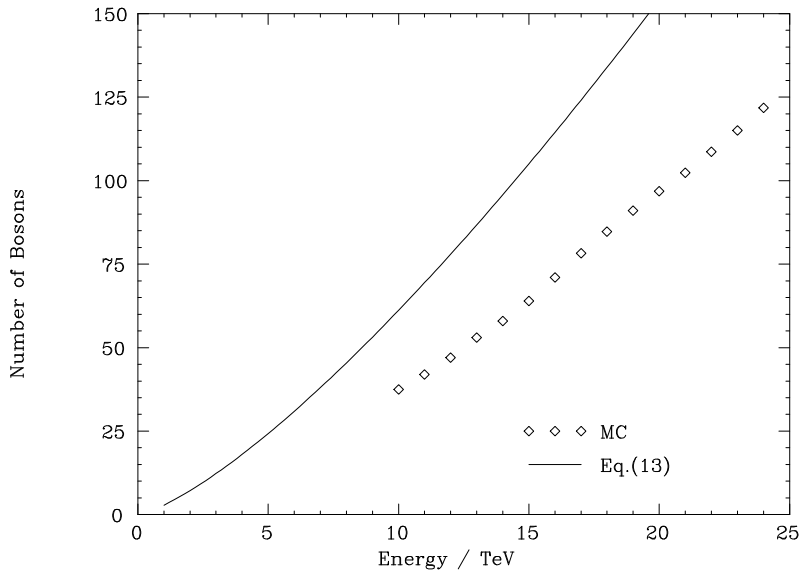}
  \caption[]{Mean number of bosons as a function of energy. Diamonds \LOME\
prediction, solid line Eq.~(\ref{eq:mul}).}
\label{fig:csnben} 
\end{figure}

This procedure also yields the mean boson multiplicity
$\bar{n}_B$ as a function of energy, as
\begin{equation}
\bar{n}_B = \frac{\sum_{n_B} n_B
\hat{\sigma}_{LO}\left( n_B,{\cal C}_f\right)}{
\sum_{n_B} \hat{\sigma}_{LO}\left( n_B,{\cal C}_f\right)} .
\end{equation}
A plot of $\bar{n}_B$ is shown in
Fig.~\ref{fig:csnben}, along with the estimate from the
`holy grail' function $F$, which is shown as a solid line
on the same plot.
The difference is due to the reduced phase space available
when one considers massive electroweak bosons.

At a given multiplicity $n_B$, integration over all possible
SU(2) configurations of the instanton produces a non-trivial
momentum structure in the final state, as noted 
by Espinosa \cite{es90}. This structure contains terms of
the form $\left( p.q\right)$ and 
$\left(\epsilon_{\mu\nu\alpha\beta}p_{\mu}q_{\nu}k_{\alpha}l_{\beta}\right)$,
where
$p$, $q$, $k$ and $l$ are particle momenta and polarisation vectors.
The large number of particles from the instanton interaction 
means that there will be many $\left(\sim\left[ 4 n_G + 2 n_B + 
n_H - 1 \right] !! 
\right)$\footnote{We use the double factorial notation, whereby 
$n!! = n(n-2)(n-4)...3.1$ for odd $n$.} such terms~\cite{gi94b}.
Whilst calculation of all these terms is not feasible, it is reasonable
to assume that their overall effect will be to produce 
an approximately uniform distribution in solid angle.
However, these terms could 
modify the energy spectra of the outgoing particles.
For example, the \LOME\ expression (\ref{eq:bvics})
was obtained without performing
any SU(2) averaging, and in
this case the bosons each have a factor of
$\left( 4E^2-m_W^2\right) /m_W^2$. This factor favours
configurations with higher boson energy. 
Such a factor has not been included in the simulations when either the
$n_B = 1/\alpha_W$ or \LOME\ prescriptions were used.

The averaging over SU(2) configurations also requires that the number
of Higgs bosons involved in the interaction is even. This can be
identified with the observation that the integration over all SU(2)
configurations of an odd number of rotation matrices is 
identically zero. Recall that the number of Higgs 
particles $\bar n_H$ is
expected to be small from~(\ref{eq:mulH}).
We model the number of Higgs particles at a given boson multiplicity
$n_B$ using the relative sizes of the \LOME\ estimate~(\ref{eq:bvics}) 
as a function of $n_H$. 

\subsection{Fermion final states}\label{sec:ferfin}

The structure of instanton-induced \BLNV\ involves
three quarks and a lepton from each fermion generation. 
This structure arises because the instanton determinant includes
a zero mode for each fermion doublet; there is a quark doublet for
each of the three colours of \QCD.
The integral over all SU(2) configurations ensures overall charge
conservation, and any difference in total charge between the incoming
and outgoing fermions is balanced by the relative numbers of
charged bosons.

In order to generate the fermion configuration, one member of each
doublet has to be included in the interaction. The incoming quarks
from the protons determine two of the three quark doublets for the
lightest generation, leaving the other seven antiquark and three
antilepton doublets to be chosen.

Our simulations use a simple phase space model to perform the fermion
selections. The calculation of the phase space for $n$ particles,
$V_n(s)$, by the saddle point method leads, via a Laplace transform,
to the expression
\begin{equation}\label{eq:ffsps}
V_n(s) = \frac{1}{4\pi^2 i\sqrt s}
\int_{c-i\infty}^{c+i\infty} d\beta\, \beta^2
I_1\left(\beta\sqrt{s}\right)\Phi\left(\beta\right)
\end{equation}
where
\begin{equation}\label{eq:fspst}
\Phi\left( \beta\right) = \prod_{i=1}^{n}\phi_i\left(\beta\right)
\end{equation}
is the product of the individual contributions of each particle to
the overall phase space, and $I_1$ is the entire modified Bessel
function. The 
value of $c$ is chosen so that all
singularities of the integrand lie to the left of the contour in the
complex \be\ plane.  
The contribution of a particle of mass $m_i$ is
\begin{equation}
\phi_i =  \frac{2\pi m_i}{\beta} K_{1}\left( m_i\beta\right)
\end{equation}
where $K_1$ is the singular modified Bessel function, and 
in the massless case is simply
\begin{equation}
\phi_i = \frac{2\pi}{{\beta}^2}\;.
\end{equation}
The integral (\ref{eq:ffsps}) can be
performed in the saddle point approximation.
Note that
the limiting values of the saddle point, {\barb}, for 
the cases of all $n$ particles
in the extreme relativistic and non-relativistic limits are
\begin{eqnarray}
\barb_{ER} & = & \frac{2n - 3/2}{\sqrt{s}} \;, \nonumber \\
\barb_{NR} & = & 
\frac{3}{2}\frac{\left(n-1\right)}{\sqrt{s} - \Sigma m_i}
\end{eqnarray}
respectively. 

Neglecting the effect on \barb\ of choosing 
between two particles in a doublet
with different masses, we can estimate the relative probability of
picking a particle of mass $m_1$ as opposed to one of mass $m_2$ as
\begin{eqnarray}
p & = & \frac{\phi_1}{\phi_1 + \phi_2} \nonumber \\
 & = & \left( 1 + \frac{m_2 K_1\left(m_2\barb\right)}{
m_1 K_1\left(m_1\barb\right)} \right)^{-1}
\end{eqnarray}
for two massive particles, and by
\begin{equation}
p = \left( 1 + m_2\barb K_1\left(m_2\barb\right) \right)^{-1}
\end{equation}
for a doublet with $m_1 = 0$. 

\begin{figure}[tb]
  \vspace*{8.2cm}
  \includegraphics{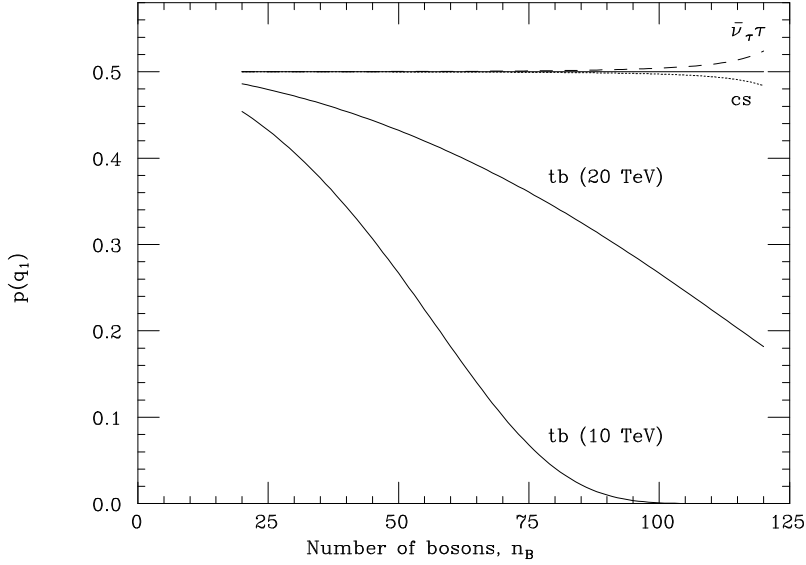}
  \caption[]{Relative probability of emission for fermion doublets.
Energy = 20~TeV except for the lower $tb$ line.}
\label{fig:tbrelp} 
\end{figure}

We find that, with the exception of the \tbar\bbar\ doublet, the
effect of mass on the probability $p$ is negligible. For
the heavy quark case, the relative probability of
\tbar\ compared to \bbar\ production is 
reduced as the boson multiplicity increases, as shown in
Fig.~\ref{fig:tbrelp}. It is evident that
the only significant effect occurs at large boson
multiplicity, where the phase space is sensitive to mass effects.

The fermion configurations are described in terms of the gauge eigenstates
of the electroweak theory. It is the mass eigenstates of the 
quark fields that are observed, and  the gauge eigenstates
of the $I_{3}^W= -${\small{$\frac{1}{2}$}} quarks have to be projected
onto the mass ones. The states are related by the CKM matrix $U$ by
\begin{equation}
|{\mathrm{mass}}\rangle = U |{\mathrm{gauge}}\rangle\; .
\end{equation}

\subsection{Hadronisation of B violating final states}
\label{sec:colint}

The objective of a \MC\ event generator is to generate
complete final states in all the detail that could in principle
be observed by a perfect detector. In the case of colourless
produced particles (electroweak and Higgs bosons, and leptons)
it is a reasonable approximation to suppose that they interact
with the detector or decay independently, apart from possible
spin correlations that have not yet been computed but are probably
of secondary importance.  For coloured particles (quarks and gluons),
however, we know that confinement plays a r\^ole that 
cannot be neglected,
giving rise to hadronic jets whose properties are only indirectly
related to those of the primary partons.  One of the main successes
of modern event generators 
\cite{hw92,herwigphys,js84,js87} has been the surprisingly
accurate representation of jet properties that they give when applied
to conventional processes such as $\ee$ annihilation. It is therefore
natural to take over the detailed machinery of jet fragmentation in
such generators and apply it, unchanged as far as possible, to the
unconventional process of baryon number violation.

According to the viewpoint adopted in all the widely-used \QCD\ event
generators, the process of jet production and fragmentation takes
place in three distinct phases.  First, on the very shortest relevant
timescale (the electroweak scale in this case), quasi-free partons are
produced with the distributions prescribed by the hard process matrix
elements. Next, these primary partons give rise to {\em parton showers}
via successive gluon bremsstrahlung and, less commonly, quark-antiquark
pair production. The showers can be followed perturbatively until the
timescale approaches the hadronic scale $1/\lqcd$, when the
running \QCD\ coupling $\as$ becomes large. At this stage the shower
development is terminated by imposing a parton virtuality cutoff
$Q_0 > \lqcd$.

The final phase of jet fragmentation, in which the showers at the cutoff
scale $Q_0$ are converted into hadrons, is called {\em hadronization}.
This process is not well understood at a fundamental level
but various models have been developed which can describe
it in some detail. In the \HW\ \MC\ program\cite{hw92,herwigphys},
a cluster model \cite{clus} is used: all remaining gluons in
the showers are split non-perturbatively
into quark-antiquark pairs and colourless mesonic clusters form in the
resulting cloud of quarks and antiquarks. The observed hadrons come
from cluster decay according to a simple phase-space model.  In the
\JS\ program \cite{js84,js87} a string 
model \cite{string} is used: instead
of clusters, more extended colour-singlet objects called strings are
formed and these fragment sequentially into hadrons.

In the generation of the parton showers, and also for the formation of
clusters or strings during hadronization, it is essential to keep track
of the colour structure of the process, at least to leading order in
$1/N$ where $N$ is the number of colours. An effect of special
importance is the colour coherence of soft gluon emission by
the primary partons, which sets the limits on parton showering
via angular ordering \cite{ang81a,ang81b}.  We first 
review how this works
for conventional baryon-number conserving processes,\footnote{For
a more extensive review, see for example Ref.~\cite{QCDbook}.}
and then show how it applies equally well in the presence of
baryon number violation.

\begin{figure}[tb]
  \vspace*{6.9cm}
  \includegraphics{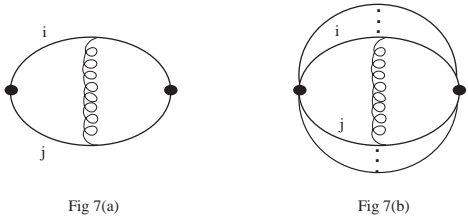}
  \caption[]{Soft gluon coherence: (a) emission from
a $q\bar{q}$ pair, (b) emission contribution from two quarks after a
\BLNV\ event.}
\label{fig:BVI}                                                         
\end{figure}

Consider for example the diagram in Fig.~{\ref{fig:BVI}}a,
representing a contribution to the matrix element
squared for emission of a gluon of momentum $q$ from a
colour-singlet quark-antiquark pair $ij$.  In Feynman gauge (neglecting
quark masses) this is the only contribution, and in the soft limit
it gives a spin- and colour-averaged emission probability of the form
\begin{eqnarray}
\label{eq:qqbarg}
dP_{ij} &=& \frac{g^2}{N}\mbox{Tr}(t^a t^a)
\frac{p_i\cdot p_j}{p_i\cdot q\; p_j\cdot q}\,\frac{d^3q}{2(2\pi)^3q}
\nonumber \\
&=& C_F \frac{\as}{\pi}\frac{dq}{q} \frac{d\Omega}{4\pi} W_{ij}
\end{eqnarray}
where $t^a$ are the colour matrices in the fundamental representation,
$C_F = \mbox{Tr}(t^a t^a)/N = (N^2-1)/2N$ and the radiation function
$W_{ij}$ is given by
\begin{equation}
\label{eq:Wij}
W_{ij}=
 \frac{(1-\cos\theta_{ij})}{(1-\cos\theta_i)(1-\cos\theta_j)}
\equiv \frac{\xi_{ij}}{\xi_i \xi_j}\;,
\end{equation}
$\theta_{ij}$ being the angle between lines $i$ and $j$ and $\theta_i$
 ($\theta_j$) the angle of emission of the gluon with respect to
line $i$ ($j$).

The way in which the radiation function (\ref{eq:Wij}) leads to angular
ordering can be seen as follows. We write
\begin{equation}
\label{eq:Wijsep}
W_{ij} = W^i_{ij} + W^j_{ij}
\end{equation}
where
\begin{equation}
\label{eq:Wiji}
W^i_{ij}
= \frac{1}{2\xi_i}\left(1+\frac{\xi_{ij}-\xi_i}{\xi_j}\right)\;.
\end{equation}
Then $W^i_{ij}$ and $W^j_{ij}$ contain the leading collinear singularity
as $\theta_i \to 0$ and $\theta_j \to 0$ respectively.  Furthermore
if we average $W^i_{ij}$ over $\phi_i$, the azimuthal angle of emission
with respect to line $i$, we find precisely
\begin{equation}
\label{eq:Wijiav}
\VEV{W^i_{ij}}_{\phi_i}
= \frac{1}{\xi_i}\Theta(\xi_{ij}-\xi_i)\;.
\end{equation}
That is, after azimuthal averaging, $W^i_{ij}$ represents emission inside
a cone centred on line $i$ with half-angle $\theta_{ij}$. Similarly
$W^j_{ij}$ represents emission in a cone centred on line $j$ with
the same half-angle.  Outside these cones, the radiation averages to
zero, which is what is meant by angular ordering. In the `coherent
parton shower' approximation used in \MC\ simulations \cite{mw84},
the radiation function is set equal to zero outside the angular-ordered
region. This gives the correct results for quantities that are
azimuthally averaged, such as multiplicity and energy distributions.
The approximation is also found in practice to be quite accurate for
global quantities that are in principle sensitive to azimuths,
such as event shapes, because the contributions from
angular-disordered regions tend to cancel and are in any
case numerically small, such regions being far from all collinear
singularities.  A similar analysis can be performed for soft gluon
emission from heavy quark systems: here the collinear singularities
are screened but angular-ordered regions are still enhanced \cite{heavy}.

The extension of the above treatment to baryon-number violating processes
poses no special problems since colour, together with other generators
of local gauge symmetries, remains absolutely conserved in the hard
process.  We consider the soft gluon emission contribution in which
the gluon connects lines $i$ and $j$ emerging from a baryon-number
violating vertex, Fig.~{\ref{fig:BVI}b}. For $N$ colours, the 
vertex couples $N$
quarks via the invariant tensor $\epsilon_{ijk\ldots}$. The emission
probability becomes
\begin{equation}
\label{eq:bvig}
dP_{ij} = C_B \frac{\as}{\pi}\frac{dq}{q}
\frac{d\Omega}{4\pi} W_{ij}
\end{equation}
where the colour factor is
\begin{equation}
\label{eq:CB}
C_B = -\frac{\epsilon_{ijk\ldots}\epsilon_{i'j'k\ldots}t^a_{ii'}t^a_{jj'}}
{\epsilon_{ijk\ldots}\epsilon_{ijk\ldots}}
=\frac{\mbox{Tr}(t^a t^a)}{N(N-1)} = \frac{C_F}{N-1}\; .
\end{equation}

If we now make the decomposition (\ref{eq:Wijsep}) of the radiation
function, we see that the full contribution with the leading collinear
singularity along the direction of line $i$ is given by Eq.~(\ref{eq:qqbarg})
with the replacement
\begin{equation}
\label{eq:bviji}
W_{ij} \to \frac{1}{N-1}\sum_{j\neq i} W^i_{ij} \; .
\end{equation}
That is, we should simply average the contribution with respect to all
other lines emerging from (or entering) the same baryon-number violating
vertex. In the coherent parton shower approximation, this means that we
should choose the cone which limits emission from line $i$ to have half-angle
$\theta_{ij}$ where $j$ is chosen at random from all other lines at the
same vertex.  Within this approximation, the parton showers from each
line then develop independently.

As already mentioned, in the cluster hadronization model adopted in
the \HW\ program \cite{hw92} all gluons remaining at the end of
parton showering are split into quark-antiquark pairs. Colour
singlet $q\bar q$ combinations are then used to build clusters.
This is actually done in the leading-$1/N$ approximation, which
greatly simplifies matters since each external colour line in any
diagram is then uniquely connected to an external anticolour line
with which it forms a singlet. A similar approximation is used
in \JS\ \cite{js84,js87} to decide which quarks and gluons should
be connected by string segments. For baryon-number
violating processes essentially the same procedure can be adopted:
all connected external colour-anticolour pairs are first used
to form singlet mesonic clusters or strings, leaving sets of $N$ colour
or anticolour lines connected to each other via the same hard vertex,
which form singlet baryonic or antibaryonic clusters/strings. Any excess
baryons or antibaryons naturally appear amongst the decay products
of the latter.

It follows from the above hadronization mechanism that the 
excess baryons in a B-violating final state tend to be
amongst the lower -momentum jet fragmentation products,
rather than the leading particles which carry the bulk of the
jet momentum. An excess baryon can be a leading particle only
if the corresponding excess quark does not emit any gluons above
the hadronization cutoff $Q_0$. The probability of this is
suppressed by a Sudakov form factor, which decreases faster than any
inverse power of the hard process scale.

\subsection{Backgrounds to B and L violating processes}\label{sec:blback}

We consider three possible backgrounds to \BLNV\ processes:
non-perturbative multi--W(Z) production, multiple electroweak
boson production by perturbative standard model processes, and
\QCD\ multijet production.

Non-perturbative multi--W production
is expected to occur at high multiplicities, \nw$\sim 1/\alpha_W$, 
with an energy scale $\sim m_W/\alpha_W$.\ A detailed discussion,
based on the argument that at high energies electroweak theory has
a large, essentially constant total inelastic cross section,
may be found in ref.~\cite{ri91b}.
The fermion structure for 
multi-W production without \BLNV\ is
\begin{equation}
\label{eq:mwqeqn}
q_1~+~q_2\rightarrow ~q_3~+~q_4
~+~n_B~\mbox{W(Z)}~+~n_H~\mbox{H},
\end{equation}
giving a signature similar to the \BLNV\ case
considered previously (\ref{eq:anoproc}), without the 
primary fermions.
It is expected that these processes
will have a similar threshold behaviour and cross section
to the \BLNV\ ones. We have therefore modelled this background
in the same manner as the \BLNV\ signal.
In principle, there are also processes initiated by initial-state
antiquarks, but again, because of the high threshold
the $qq$ combination dominates (see Fig.~\ref{fig:qqconv}).
For simplicity, we exclude the other
incoming combinations, as in the \BLNV\ case.
                                                
Perturbative multiple production of electroweak bosons by standard model
processes has been estimated in refs.~\cite{ba92,dm93}.  The dominant mechanism
is expected to be multiple top quark production: each $t$ decays into a
$W$ boson and a $b$-quark jet.  At 
40 TeV energy the expected $t\bar t$
cross section is $3\times 10^6$ fb while that for $t\bar t t\bar t$ is expected
to be about 400 fb.  In contrast, the predicted direct $4W$ cross section
is only around 6 fb, which is comparable to the rate via Higgs bosons
decaying into $WW$. Clearly, in order for the production of much
larger numbers of $W$ bosons to be observable, a breakdown of
perturbation theory is required, as assumed above.

Even though each jet from $W$ decay is in principle well defined, for
high $W$ multiplicities the number of possible dijet combinations becomes
too high for the $W$'s to be reconstructed. At very high multiplicities,
the jets will be too close together to be resolved.  A typical jet cone
size, R\footnote{
$R$ is defined as 
$R = \sqrt{\Delta\eta^2\ +\ \Delta
\phi^2}$, $\eta = -{\mathrm{ln\ tan}} \theta/2$ is the pseudorapidity
and $\phi$ is the azimuth, both with respect to the beam
direction.}, of 0.7 means that 
within a rapidity limit of $|\eta|< 3$\ only 
$12\pi/(\pi 0.7^2) = 24$ jets
can be accommodated without overlapping. Thus, for hadronic $W$ decays, we
have to worry also about pure \QCD\ multi-jet background.

Perturbative \QCD\ multi-jet production has been studied most in the process
of $W + n$ jet hadroproduction \cite{gi92,be91}, since this is an
important background for top quark searches. The predicted $W+2$
jet cross section at 40 TeV for jet
$p_T>50$ GeV, $\Delta R_{jj} > 0.7$ and $|\eta_j|< 3$ is about
$2\times 10^6$ fb.  Tree-graph calculations for 3 and 4 jets suggest
a fall in the cross section by a factor of 2-3 for each additional
jet, as long as phase space limitations can be neglected. 

\QCD\ multi-jet production without any direct $W$ boson has been less
studied.  For the same cuts as above, the predicted 2-jet cross section
is about $4\times 10^{10}$ fb. Four and six jet production can be
estimated from ref.~\cite{ch91} at roughly $4\times 10^8$ and
$4\times 10^6$ fb, suggesting a fall of around a factor of 10
for each additional jet.
However, these estimates are based on multi-gluon amplitudes, and
the $W + n$ jet studies suggest that multi-quark amplitudes play
a significant role at high multiplicities. This may explain the smaller
rate of decrease seen in those studies.  The \MC\ estimates
of ref.~\cite{od90} are indeed somewhat higher, but this may be due
to a jet definition in terms of invariant mass, $m_{jj}>6$ GeV,
which corresponds to a very narrow $\Delta R_{jj} > 0.1$ at $p_T=50$ GeV.
A fall by a factor between 2 and 10 per additional jet
would imply, for example, a perturbative
30 jet cross section of $10^{-9\pm 9}$ fb. 

In a recent paper \cite{go93}, approximate expressions for high-order
multigluon tree amplitudes are used to derive the following estimate
for the \QCD\ $n$-jet cross section
\begin{equation}
\sigma_n \sim \frac{1}{s} [z(n,\Delta)]^{n-2}
\end{equation}
where $s$ is the hard scattering energy scale and $\Delta = m^2_{jj,min}/s$.
For $\ln(1/\Delta)\gg n$, the suppression factor $z(n,\Delta)$ has the
form
\begin{equation}
z(n,\Delta)\sim \frac{N_c\alpha_S}{2\pi\sqrt{12}}\ln^2(1/\Delta)
\end{equation}
For the jet definitions used above, taking ${\sqrt{s}}\sim$\ 10 TeV gives
$\ln(1/\Delta) \gtap 10$ and $z>1$.  Thus the approach of
ref.~\cite{go93} implies a breakdown of \QCD\ perturbation theory.
Indeed, the instanton approach adopted here for electroweak physics
suggests that in \QCD\ such a breakdown would be expected for
$n\sim 1/\as$, with an energy scale of $\lqcd/\as$~\cite{ri91b}.

For studies of lepton number violation, the lepton content of the 
background is relevant. 
Multi--W production and decay would yield many hard charged leptons
(about 0.25 per $W$) which would be confused with those from the
primary process. In \QCD\ multijet production, from
\MC\ studies of 2-jet events
with jet $p_T>50$ GeV at 40 TeV, the predicted number of charged
leptons per jet was found to be 0.07, mostly from charm decay.
Thus 30 jets would provide only 2 charged leptons on average,
and these would mostly appear at low transverse momentum.

In conclusion, the cross sections for backgrounds to \BLNV\ are
as uncertain as that for the signal. However, if a signal exists, then
B and L conserving multi--W processes are likely to occur at a similar rate,
and constitute the background most difficult to distinguish from
the signal. Consequently we shall concentrate on this background in
the \MC\ studies in the next section.


\section{Monte Carlo results}
\label{sec:MCres}

In this section we describe the results obtained by performing \MC\ 
simulations of \BLNV\ processes. The events under consideration
are very distinctive, owing to the high number of particles involved. 
In particular, the large number of leptons present and the high total
transverse energy $E_T$ will be useful experimental criteria for
isolating multi--W and \BLNV\ events.

Four cases were studied to investigate 
the observability of B and/or L violation.
The configurations of the parameters {\nw} and  $\sqrt{\hat{s}_0}$
used are listed in Table \ref{tab:sims}, along with approximate
values for the energy fraction $f_W$ and the mean energy given
by $\bar{E}$. This 
range of $f_W$ and \nw\ allows useful
comparisons to be made.
The simulations were performed using an event generator,
\HERBVI~\cite{gi94a}, which incorporates the parametrizations and
assumptions described in the previous section.
\HERBVI\ is a package designed as an extension to
the \MC\ program \HW, which performs the hadronisation of the
final state as described in Sect.~\ref{sec:colint}. Some
further details are given in the appendix.
\begin{table}
\begin{center}
\begin{tabular}{|rc|c|c|c|r|} \hline
\multicolumn{6}{|c|}{Simulations performed} \\ \hline \hline
\multicolumn{2}{|c|}{Energy (TeV)}& 
$n_B$ estimate & $\sqrt{\hat{s}_0}$\ (TeV) & $f_W$ 
& $\bar E$ \\ \hline
17 & & $1/\alpha_W$ & 5 &1.8 & 60 \\
40 & (a) & $1/\alpha_W$ & 18 & 7.4 & 450 \\
40 & (b) & \LOME & 18 & 2.0 & 80 \\
200 & & $1/\alpha_W$ & 18 & 9.3 & 580 \\ \hline
\end{tabular}
\end{center}
\caption{\MC\ simulations performed. All simulations
contained a $10^{4}$ event sample.}\label{tab:sims}      
\end{table}

The lack
of more detailed knowledge of the matrix element means that we have
concentrated on the gross properties of the events, such as the
excess of antileptons over leptons, as opposed to detailed analysis
of momentum spectra or
rapidity distributions.
In each of the four cases
studied $10^4$ simulated events were generated, with
$m_{top} = 140$\ GeV and $m_{Higgs} = 300$ GeV.

In order to assess the detectability of \BLNV, we
have conceived a hypothetical detector loosely based on the proposals
for detectors at the \LHC \cite{Atlas,CMS}. 
The simulations were performed with
the following acceptance criteria:

\begin{itemize}
\item A minimum of four identified charged leptons of the same 
family ($e$\ or $\mu$) for
an event to be accepted
\item Pseudo--rapidity coverage up to $|\eta| = 3$
\item Minimum particle transverse momentum $p_t$ of 10 GeV  
\item Electron and muon identification efficiency 95\%.
\end{itemize}

In addition, we 
define isolated leptons and photons in the same way as
the \ATLAS\ \LHC\ detector proposal \cite{Atlas}.
A particle is considered to be isolated if there is
a total of less than 12 GeV energy deposited 
in the detector calorimetery
around it within a radius
$R < 0.2$.

\begin{figure}[tb]
  \vspace*{8.2cm}
  \includegraphics{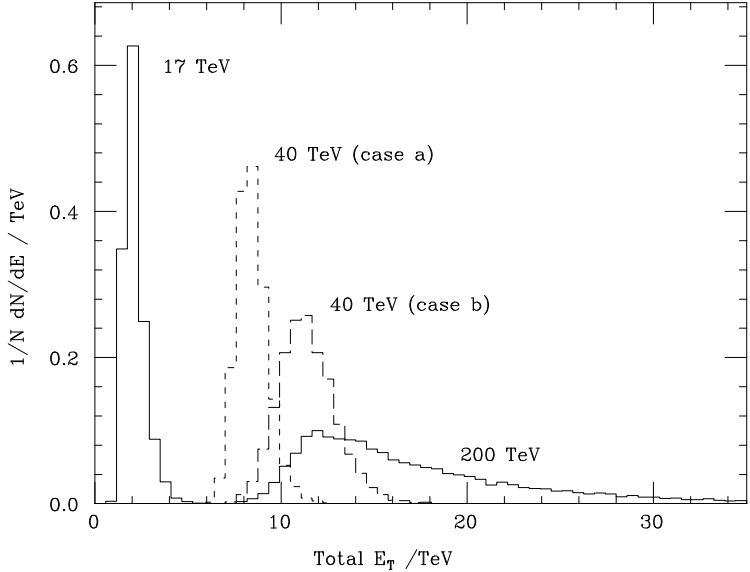}
  \caption[]{Total transverse energy distribution.}
\label{fig:etrans}
\end{figure}

No cut on $E_T$ has been made in the simulations. In Fig.~\ref{fig:etrans}
the distribution of $E_T$ for the four simulations has been plotted. Here
$E_T$ is defined as the sum of transverse 
energy $\left( m^2+p_T^2\right)^{1/2}$ for 
all {\em detected} particles.
A minimum $E_T$ of $5$ TeV ( 1 TeV for the 17 TeV simulation) will include all
the events generated here, yet reject virtually all other types of events.
Obviously, this cut would have to be `tuned' in an experimental analysis,
but combined with the multiple lepton cut the background from any
non--multi--W standard model process should be essentially zero.

\subsection{Boson Spectra}

With the possible exception of photons, the study of bosons from
instanton-induced processes will be difficult. It may be possible to
reconstruct the leptonic decays of \Z's;
for \W\ decays
this will not be possible due to the large missing momentum from the
neutrinos. The identification of jets, and hence the reconstruction of
hadronic \W\ and \Z\ and decays, will also be difficult because of the
large number of overlapping jets.
However, the energy spectra of the bosons
produced in the initial interaction 
are an important aspect of the process
since they reflect the characteristic size of the
interaction region $\sim 1/m_W$.

\begin{figure}[tb]
  \vspace*{6.9cm}
  \includegraphics{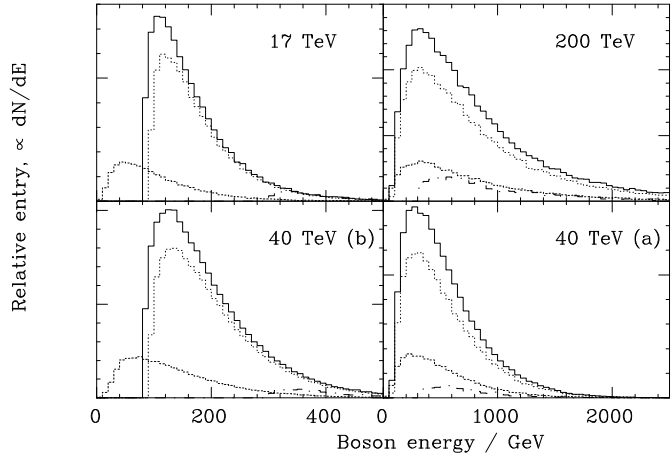}
  \caption[]{Boson energy spectra. Area under each curve is the average
number of bosons by type per event. Solid $W^{\pm}$, dotted
$Z^{0}$, dashed photons, dot-dashed Higgs ($\times 10$). }
\label{fig:bosspec}
\end{figure}

The boson energy spectra for the simulations are plotted in
Fig.~{\ref{fig:bosspec}}. The distributions have been weighted so that
the total area under each graph is equal to the average number of
each type of boson produced in the interaction, with the exception
of the Higgs boson results which have been increased by an 
factor of ten.
The much lower average boson energy of the $f_W\sim 2$ simulations
compared to the other two can clearly be seen. 

\begin{figure}[tb]
  \vspace*{8.2cm}
  \includegraphics{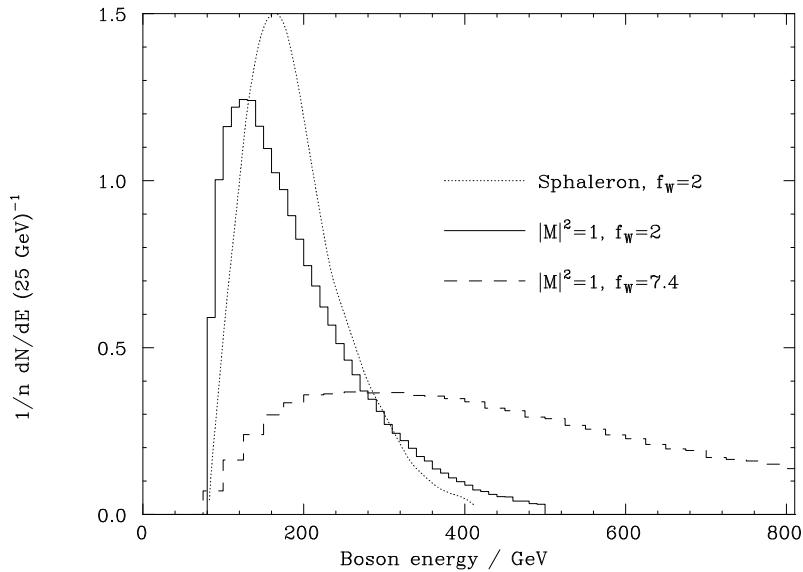}
  \caption[]{Comparison of boson energy spectra with sphaleron
decay spectrum.}
\label{fig:sphalcom}
\end{figure}

The $W^{\pm}$ distributions 
from these plots are shown again in Fig.~\ref{fig:sphalcom}
along with the sphaleron decay spectra~\cite{za92} for comparison.
The sphaleron decay corresponds to $f_W = 2$. It can be seen that
the distribution differs slightly from that of the $f_W = 2$ simulation.
The calculation of the sphaleron decay products involves the solving
of the non-linear field equations. Therefore the spectra of the 
outgoing particles from this decay include the effects of final
state corrections (but not of the integration over orientations of
the classical background field). The similarity of the curves 
shows that the assumption of a flat distribution of outgoing 
particles in phase space
is a reasonable first approximation. Note however that this assumption
yields more particles in the
high momentum tail of the distribution. 

\begin{figure}[tb]
  \vspace*{5.0cm}
  \includegraphics{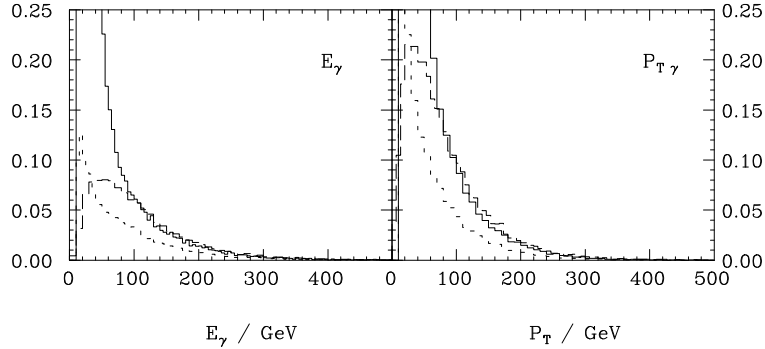}
  \caption[]{Photon energy spectra for the 17 TeV simulation. Solid line all
photons, dashed photons from primary interaction, short-dashed 
isolated photons.}
\label{fig:phot17}
\end{figure}

\begin{figure}[tb]
  \vspace*{5.0cm}
  \includegraphics{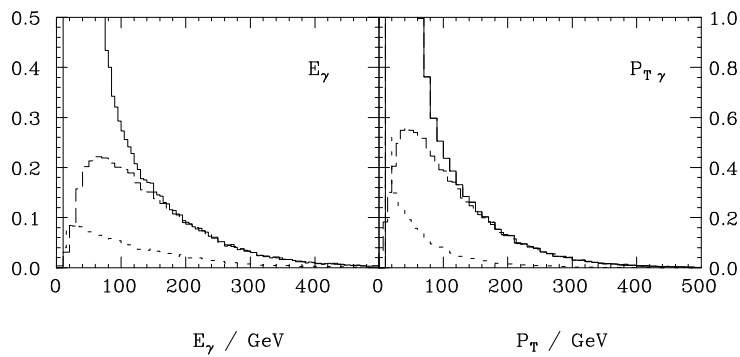}
  \caption[]{Photon energy spectra for the 40 TeV simulation, using
the leading order matrix element estimate for {\nw}. Solid line all
photons, dashed photons from primary interaction, short-dashed 
isolated photons.}
\label{fig:phot40b}
\end{figure}

Direct photons from the primary interaction, which
have not been considered before, may provide the best
opportunity
to investigate the bosonic
spectra of the instanton interaction. 
In Fig.~\ref{fig:phot17} the
photon $E$ and $p_t$ spectra from the 17 TeV simulation are
plotted. Typically $\sim 60$ photons are produced per event,
the majority of which are from pion decay and at the low end of the
energy spectrum. The tails of the distributions largely comprise
direct photons from the primary interaction. The average number of
photons produced in this way 
is $n_B\,\sin^2 \theta_W / 3$, which
corresponds to $\sim 2.2$ for 
the $n_B\sim 1/\alpha_W$ and $\sim 7.7$ for the
\LOME\ estimate respectively.
The short-dashed line is the spectrum of photons after the
isolation cut had been imposed. For this simulation, the
domination of the photon spectra by primary photons
for $E > 100$ GeV and $p_t > 100$ GeV means that
measurement of photons in this region would provide information on
the \BLNV\ process structure. Similar conclusions can be drawn
from the other two $n_B\sim 1/\alpha_W$ cases, but in the
40 TeV simulation using the \LOME\ {\nw}\ estimate,
shown in Fig.~\ref{fig:phot40b}, there are very few isolated
photons. This is because of the larger number of particles in the 
detector, which makes finding an isolated particle less likely.


The isolation cut removes the lower energy particles, leaving
the tails which are
mainly composed of photons from the primary interaction. The 
study of these tails will provide useful information about the underlying
process.
Any structure from the matrix element of the instanton process will
show up in this part of the spectrum. 
Therefore measurements of  
photon spectra will be important in any experimental study of
\BLNV\ or multi--W phenomena  in general.
Good calorimetry, especially if a `tighter' definition of isolation
is possible, would aid this task considerably.

\subsection{Baryon number violation}\label{sec:bviolr}

The high energy threshold of the \BLNV\ processes we are
studying, along with the `central' nature of the events, means
that the events are characterised by a large number of particles,
$\cal O\left( 10^{3-4}\right)$, within
the detector.
Nevertheless, many final state particles
escape detection,
either because they do not enter
the solid angle covered by the detector, or because their energies
are so low that they cannot be reliably detected.
Consequently, any verification of \BLNV\ will have to
concentrate on averaging quantum numbers over a number of events,
with the aim of demonstrating an asymmetry. We therefore focus our
analysis on determining quantities that can be used to demonstrate
such asymmetries, and on the number of events that will be required.

\begin{figure}[tb]
  \vspace*{5.9cm}
  \includegraphics{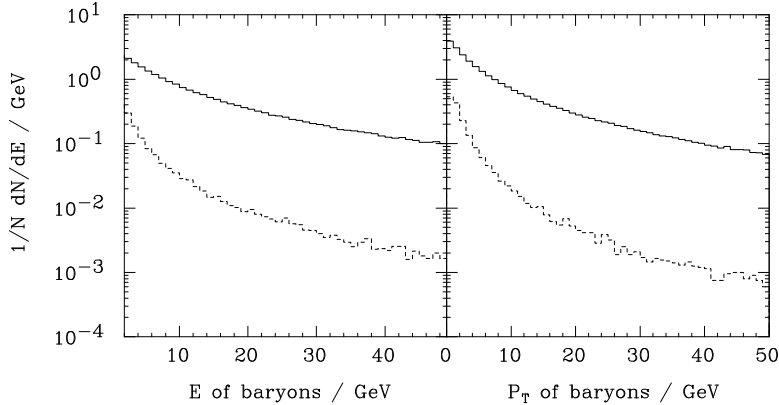}
  \caption[]{Energy and transverse momentum spectra for baryons and
antibaryons. Solid line, all baryons and antibaryons, dashed  
antibaryons from \BLNV\ vertex only.}
 \label{fig:barept} 
\end{figure}

The large number of particles present in the detector makes the
identification of baryons difficult. This is particularly true
for any identification technique based upon the reconstruction
of the masses of unstable baryons.
Furthermore, the excess antibaryons from the
\BLNV\ process are at low energies and transverse
momentum, as can be seen in Fig.~\ref{fig:barept}. The task of
separating these antibaryons from the others produced in the
interaction is effectively impossible.

\begin{figure}[tb]
  \vspace*{8.2cm}
  \includegraphics{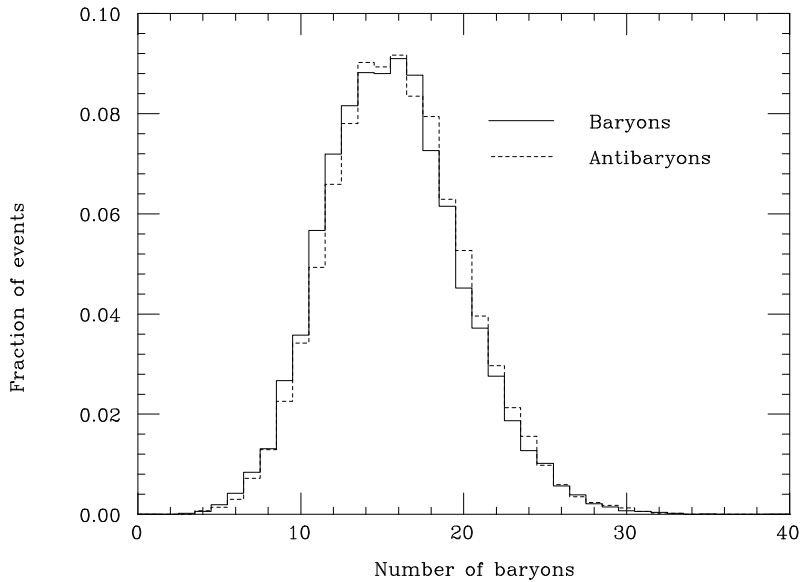}
  \caption[]{Number of baryons and antibaryons observed in detector
per event. Solid line baryons, dotted line 
antibaryons.}                      
\label{fig:barnum}
\end{figure}

We may 
demonstrate the difficulty of verifying B violation
by making an estimate of the number of events required for 
observing B violation with a 95\%
($3\sigma$) confidence level, assuming for the moment
perfect baryon identification 
within the detector solid angle. 
We define the baryon number difference $D_{B}$ as
\begin{equation}
D_{B} = 
N_{B}
 - 
N_{\bar{B}}
\end{equation}
where $N_{B}$ is the number of detected baryons in a given event, and
$N_{\bar{B}}$ the corresponding number of antibaryons.
The averaging is performed separately for \BLNV\ and multi-W events.
For the 40 TeV /BLNV/ simulation using the \LOME\ estimate for
$n_B$, the distributions
of $N_{B}$ and $N_{\bar{B}}$ are plotted in
Fig.~\ref{fig:barnum}. The means of these distributions are given
in Table \ref{tab:barnum}.
\begin{table}
\begin{center}
\begin{tabular}{|c|r@{$\pm$}c|r@{$\pm$}c|} \hline
\multicolumn{5}{|c|}{Baryon number results} \\ \hline
\hline
\multicolumn{1}{|c|}{Measurement} 
& \multicolumn{2}{|c|}{\BLNV} &\multicolumn{2}{|c|}{Multi--W}
\\ \hline
$N_{B}$ & 15.7 & 0.04 & 15.8 & 0.04 \\ 
$N_{\bar{B}}$ & 16.0 & 0.04 & 15.7 & 0.04 \\
$D_{B}$ & -0.26 & 0.03 & -0.02 & 0.03 \\ \hline
\end{tabular}       
\end{center}
\caption{Total number of baryons and antibaryons,
within solid angle covered by detector. Results
from 40 TeV simulation with $n_B$ given by the \LOME\ estimate.}
\label{tab:barnum}      
\end{table}

In order to estimate the number of events required for baryon
number violation to be established at a suitable confidence
level, we compare two sets of data. The first is a mixture of
\BLNV\ and multi--W events in equal amounts, corresponding to
an equal probability for these two processes.
The second set of data is the multi--W process alone, and we
calculate the confidence level of separating the two data
sets as a function of the number of events, using 
Student's $t$ test~\cite{nrec}, as described in appendix B. 

This analysis predicts that approximately $8.10^{3}$ events are needed
to verify B violation at the 95\% confidence level. However, this is
assuming a 100\% efficiency of baryon identification, which is far from 
realistic. The number of events required for a given confidence level
is approximately proportional to $1/{\zeta^{2}}$ for an efficiency
of $\zeta$.\ Using a figure of 1\% for this efficiency, which is 
probably a very generous overestimate,\footnote{We thank M.A.\ Parker for
useful discussions on this point.} the number of events is
then $\sim 10^{8}$. By concentrating on 
types of baryons with particularly
distinctive decay modes, it may be possible to decrease this figure, but
it is unlikely that any reduction by orders of
magnitude is possible. We conclude that
the verification of B violation is effectively impossible for
the types of processes considered here.

\subsection{Lepton number violation signatures}\label{sec:lviolr}

A more promising signature for \BLNV\ is the violation of
lepton number. 
Unlike the baryons from the primary process, which tend to 
be found amongst the low-momentum debris after hadronization
of the excess antiquarks as discussed in Sect.~\ref{sec:colint}, the
excess leptons emerge directly from the primary process with
high momenta and can be identified. 
It is not possible to demonstrate L violation for a
single event, as it is not possible to detect the produced
neutrinos. 
However, it may be possible to demonstrate an asymmetry
by considering the average number of leptons and antileptons
produced in a set of events.
In this section we discuss the
possibility of observing lepton number violation using both
model-independent and model-dependent techniques. We restrict our
discussion to $e^{\pm}$ and $\mu^{\pm}$ leptons.

Simulations performed for the four cases 
listed in Table~\ref{tab:sims}
show that L violation is
much easier to demonstrate than B violation.
Therefore a more detailed analysis of the characteristics of
the leptons is warranted. Recall that
a cut on the minimum number of detected leptons
has been imposed. In order for an event to be included
in the analysis, it has to 
contain at least four electrons or four muons. The cut chosen is 
similar to those that
will have to be made at the trigger level at any high luminosity
supercollider. This cut also helps in some
cases to reduce the multi--W
background. Table \ref{tab:cutnum} summarises the number of
events passing this cut.
\begin{table}
\begin{center}
\begin{tabular}{|rc|r|r|r|r|r|r|} \hline
\multicolumn{8}{|c|}{Events passing lepton counting cuts}
\\ \hline \hline
\multicolumn{2}{|c|}{Energy} &
\multicolumn{3}{|c|}{\BLNV} & \multicolumn{3}{|c|}{Multi--W} \\ 
\cline{3-8}
\multicolumn{2}{|c|}{TeV} & Electron & Muon & Either &
Electron & Muon & Either \\ \hline
17 & & 55.8 & 55.2 & 81.1 & 38.4 & 38.0 & 63.2 \\
40 & (a) & 72.5 & 70.5 & 92.8 & 51.1 & 49.7 & 77.2 \\
40 & (b) & 99.5 & 99.6 & 100.0 & 99.3 & 99.3 & 100.0 \\
200 & & 75.0 & 73.0 & 94.0 & 51.7 & 50.9 & 77.5 \\ \hline
\end{tabular}
\end{center}
\caption{Percentage of events passing lepton count cuts.
For the 40~TeV simulations, 
case~(a) is $n_B \sim 1/\alpha_W$, case~(b) \LOME\ prescription.}
\label{tab:cutnum}      
\end{table}
 
The simplest method of verifying the non-conservation of lepton
number is to measure the quantity 
\begin{equation}
D_{l} = N_{l^{-}} - N_{l^{+}}
\end{equation}
and demonstrate that it is incompatible with lepton number
conservation. Such a measurement can be considered model
independent in the sense that it does not depend on the
details of the matrix element of the hard process, provided the
effect of the detector cuts at low $p_T$ and high rapidity can
be neglected.

One antilepton of each family is produced by the instanton 
interaction~(\ref{eq:anoproc}). 
Consequently, we expect 
that the average of
$D_{l}$ for a large enough number of \BLNV\ events 
will tend to a value of $-0.5$\ for each lepton family. 
However, there are a number of factors which modify this prediction.
The branching ratio of a W(Z) boson to a lepton-antilepton
pair is 11\%(3.3\%). Hence, for \nw\ electroweak bosons the decay
products will include on average 0.1\nw\ leptons of each type,
with a Poisson distribution. 
The average incoming charge of the
hard process is non-zero in both the \BLNV\ and multi--W cases, which
results in a non-zero average total charge for the outgoing bosons, and
consequently a non-zero value of $D_l$ even in processes which
conserve lepton number. In addition, not all the leptons
will be detected, and there will also be contributions to $D_l$ from the
decays of heavy quarks and antiquarks. The effect of these factors can
be seen in the measured values of $D_e$ and $D_{\mu}$ listed
in Table \ref{tab:lepnum}.
\begin{table}
\begin{center}
\begin{tabular}{|rc|r@{$\pm$}c|r@{$\pm$}c|r@{$\pm$}c|r@{$\pm$}c|} \hline
\multicolumn{10}{|c|}{Lepton number results} \\ \hline \hline
\multicolumn{2}{|c|}{Energy} &
\multicolumn{4}{|c|}{$D_{e}$} &
\multicolumn{4}{|c|}{$D_{\mu}$} \\ \cline{3-10}
\multicolumn{2}{|c|}{TeV} & \multicolumn{2}{|c|}{\BLNV} &
\multicolumn{2}{|c|}{Multi--W} 
& \multicolumn{2}{|c|}{\BLNV} &
\multicolumn{2}{|c|}{Multi--W} \\ \hline
17 & & -0.48 & 0.02 & -0.09 & 0.02 & -0.47 & 0.02 & -0.07 & 0.02 \\
40 & (a) & -0.47 & 0.02 & -0.09 & 0.02 & -0.49 & 0.02 & -0.11 & 0.02 \\
40 & (b) & -0.47 & 0.03 & -0.08 & 0.03 & -0.45 & 0.03 & -0.10 & 0.03 \\
200 & & -0.40 & 0.02 & -0.04 & 0.02 & -0.41 & 0.02 & -0.10 & 0.02 \\ \hline
\end{tabular}
\end{center}
\caption{Average value of lepton number differences
$D_{e}$ and $D_{\mu}$.\ Results based on $10^4$ event
simulations.}\label{tab:lepnum}      
\end{table}

\begin{figure}[tb]
  \vspace*{5.9cm}
  \includegraphics{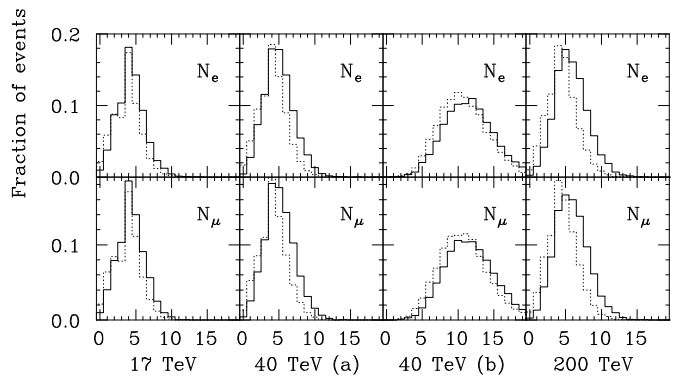}
  \caption[]{Total number of observed leptons per event. 
Solid line \BLNV\ process, dotted L conserving multi-W process.}
\label{fig:lep_tot}
\end{figure}

We first consider the general properties of the $N_{l^{\pm}}$ and
$D_{l}$ distributions before discussing the possibility of
observing lepton number violation.
The results
for $N_{l^{-}} + N_{l^{+}}$ for both muons and electrons
are plotted in Fig.~\ref{fig:lep_tot} for each of the
four simulations. The difference between the \BLNV\ and 
multi--W cases can be clearly 
seen and is due to the extra antilepton from the
primary interaction in the former case.

The number of leptons produced and detected is primarily determined
by \nw, with the main secondary effect coming from the
detector cuts and inefficiencies.
The effect of the cuts depends on 
the `centrality' of the interaction. If the threshold is large in
comparison to the total beam energy then the interaction rest frame
will tend to be nearly stationary in the detector frame. As the beam
energy increases relative to the threshold, the interaction frame
is more likely to be moving in the detector frame and consequently
more particles are likely to be lost at high rapidity.

\begin{figure}[tb]
  \vspace*{5.9cm}
  \includegraphics{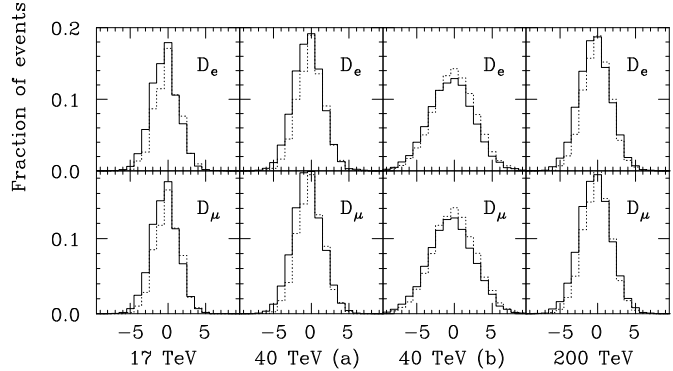}
  \caption[]{Difference in lepton number 
$D_{l}$ for muons and electrons.
Solid line \BLNV\ process, dotted L conserving multi-W process.}
 \label{fig:lep_diff}
\end{figure}

The lepton number differences $D_{l}$
for $l=e,\mu$ are plotted in Fig.~\ref{fig:lep_diff}. 
These plots show 
a similar parameter dependence to the $N_{l}$ plots.
The antileptons from
the \BLNV\ processes cause the differences between
the \BLNV\ (solid line) and multi--W (dashed line) plots.
This difference is also apparent in the average values
of the distributions, shown in Table \ref{tab:lepnum}.
In the multi--W case  $\langle D_{e}\rangle$ 
and $\langle D_{\mu}\rangle$ are close to zero. The small
non-zero values of these quantities represent the bias
from starting with incoming protons, and the other factors as
discussed above.

The mean differences $\langle D_{l}\rangle$ can be used
to demonstrate the existence of L violation.
In the same manner as the baryon case, the
number of events required to do this is estimated by comparing
an equal mixture of \BLNV\ and multi--W events to a
set of multi--W events alone. The estimates of the number
of events required to be inconsistent with the multi--W
data at the 95\%
confidence level are given in 
Table~\ref{tab:lepdiffsum}.
These estimates are written in terms of the total number of
events (\BLNV\ and multi--W combined); of the order of
$10^{3}$ events are  needed to verify the 
existence of an asymmetry.
\begin{table}
\begin{center}
\begin{tabular}{|rc|r|r|} \hline
\multicolumn{4}{|c|}{Number difference events} \\ \hline \hline
\multicolumn{2}{|c|}{Energy/TeV}& 
\multicolumn{1}{|c|}{$D_{e}$} & 
\multicolumn{1}{|c|}{$D_{\mu}$} \\ \hline 
17 & & 660 & 770 \\
40 & (a) & 890 & 670 \\
40 & (b) & 2700 & 2570 \\
200 & &900 & 1250 \\ \hline
\end{tabular}
\end{center}
\caption{Number of events for demonstrating \BLNV\ at a 95\%
confidence level by measuring lepton number differences. Case (a) at
40 TeV is 1/$\alpha_W$ number of bosons, case (b) \LOME\ prediction.}
\label{tab:lepdiffsum}      
\end{table}

We shall now consider other methods for demonstrating
lepton number violation. Our aim here is only
to investigate techniques that might aid in the verification of
lepton number violation. In particular, we consider the
possibility of enhancing the \BLNV\ event sample to improve the
determination of $D_{e}$ and $D_{\mu}$ by studying 
the momentum distribution of the leptons, and also other
indirect signatures of lepton number violation.  
We note that any analysis based on
enhancing the sample using the momenta is dependent
on the structure of the matrix element, and a quantitative 
analysis would require
more detailed modelling of the underlying processes.

It has been pointed out in \cite{fa90} that
tagging events by the sign of the highest energy lepton should
enhance the asymmetry caused by \BLNV. This is because the
most energetic $l^{-}$ in any event can only come from the
decay of a gauge boson, whereas the highest energy $l^{+}$ could
have originated from either boson decay or the primary interaction.
We define the energy
\begin{equation}
\enplu = {\mathrm{Max}}\;  E\left( l^{+}\right) 
\end{equation} 
for events with
\begin{equation}
{\mathrm{Max}}\; E\left( l^{+}\right) >
{\mathrm{Max}}\; E\left( l^{-}\right)
\end{equation}
together with the corresponding \enmin\ for 
events where the most energetic lepton is an $l^{-}$. Events with
no leptons of the type under consideration passing the
detector cuts are discarded.  

The average energy of a primary lepton from the \BLNV\ interaction, which
can only contribute to the \enplu\ distribution,  
is $\sim {\bar{E}}$, whereas a lepton from the decay of a gauge
boson, which can contribute to
both distributions, has an average energy $\sim ({\bar{E}}+m_W)/2$.
Therefore the enhancement of a sample of 
\BLNV\ events by selecting
according to the highest energy lepton will depend
on the ratio of $\bar{E}$ to $m_W$. For multi--W events, all
the leptons are produced by gauge boson decay and so there is only a small
difference between the \enplu\ and \enmin\ spectra due to
the non-zero charge of the incoming protons. 

At large \nw, it is permissible to make the approximation
$\bar m\sim m_W$. This gives
\begin{equation}
\frac{\bar{E}+m_W}{2} \sim \frac{f_W\; m_W}{2}
\end{equation}
whereas
\begin{equation}
\bar E \sim m_W{\left( f_W - 1\right)}.
\end{equation}
In the limiting case of $f_W$ close to $1$, leptons
from the \BLNV\ process
will have much less energy (close to zero) than those 
from gauge boson decay ($\sim {m_W}/2$). As $f_W$ is increased,
this imbalance is reduced. 
At $f_W \sim 2$, the average kinetic energy of
the \BLNV\ leptons will be similar to
that of leptons from gauge boson decay. 
As $f_W$ increases
above this value, the \BLNV\ leptons will tend to have
higher average kinetic energies than their boson decay
product counterparts. Therefore, as $f_W$ increases the
difference between the \enplu\ and \enmin\ spectra will
become more pronounced. 
Note that there will always be some asymmetry
between the spectra caused by the extra antilepton from
the \BLNV\ process. 
\begin{figure}[tb]
  \vspace*{6.9cm}
  \includegraphics{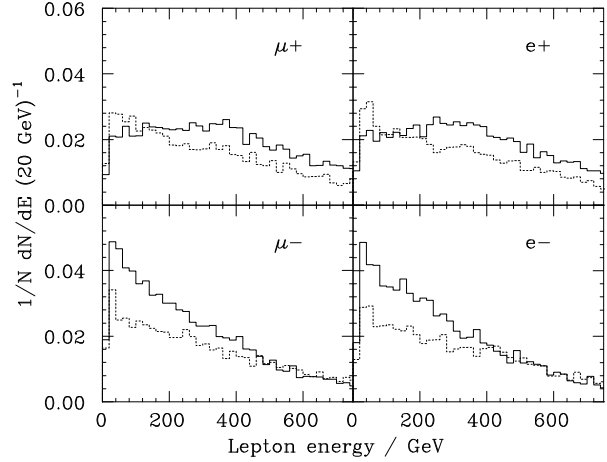}
  \caption[]{Highest energy lepton spectra 
for 40~TeV ($n_W \sim 1/\alpha_W$) simulation.
In Figs.~17-20, the solid lines represent \BLNV\ events, the dashed
lines the L-conserving background.}
\label{fig:lepen40a} 
\end{figure}

\begin{figure}[tb]
  \vspace*{6.9cm}
  \includegraphics{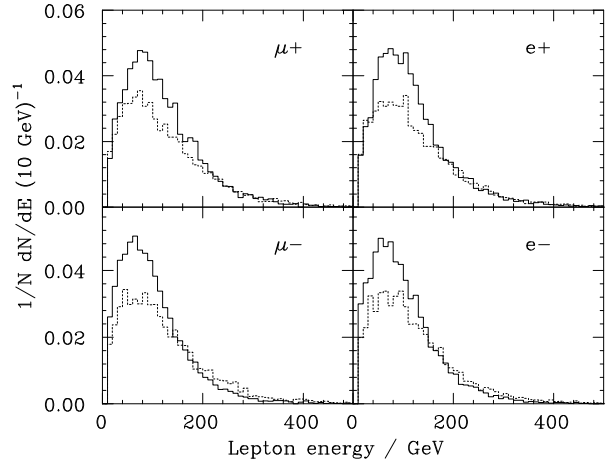}
  \caption[]{Highest energy lepton 
spectra for 40~TeV (\LOME\ prediction for {\nw}) simulation. }
\label{fig:lepen402}
\end{figure}

Using $f_W$ to classify the simulations
listed in Table~{\ref{tab:sims}}, we can divide them
into two pairs. The distributions of \enplu\ and
\enmin\ are plotted for the two 40 TeV simulations in
Figs.~\ref{fig:lepen40a}~and~\ref{fig:lepen402} 
for $f_W \sim 7.4$ ($n_B$ given by
the \LOME\ prescription) and
$f_W \sim 2$ ($n_B \sim 1/\alpha_W$) respectively. The enhancement of the
\enplu\ spectra at large $f_W$ can be clearly seen.

The spectra of Figs.~\ref{fig:lepen40a}~and~\ref{fig:lepen402} have
been normalised so that the total area under each curve is
proportional to the number of events contributing to that distribution.
By considering the relative entries to the two distributions \enplu\ and
{\enmin}, it may be possible to observe an asymmetry between the
$l^+$ and $l^-$ data. As shown in Table~\ref{tab:relheight}, a small
asymmetry is expected in the case of L conserving multi-W events, due to
the initial state charge, and a much larger one in the \BLNV\ case.
The estimates for the number of events required to demonstrate
an asymmetry, using the same mixing of \BLNV\ and multi-W events
as before, are shown in Table~\ref{tab:lepoccur}. 

\begin{table}
\begin{center}
\begin{tabular}{|rlc|c|c|} \hline
\multicolumn{5}{|c|}{Relative numbers of highest energy leptons}
\\ \hline \hline
\multicolumn{2}{|c}{Energy (TeV)}
& Lepton & {\BLNV\ events}&{Multi--W events}\\
\hline
17 & & $e^{+}$ & 58.4 & 50.2 \\
 & & $e^{-}$ & 40.4 & 46.4  \\
 & & $\mu^{+}$ & 59.7 & 49.7 \\
 & & $\mu^{-}$ & 38.9 & 47.0 \\
40 & (a) & $e^{+}$ & 64.8 & 50.1 \\
 & & $e^{-}$ & 34.8 & 47.7 \\
 & & $\mu^{+}$ & 64.7 & 50.9 \\
 & & $\mu^{-}$ & 34.8 & 47.4 \\
40 & (b) & $e^{+}$ & 55.8 & 50.6 \\
 & & $e^{-}$ & 44.2 & 49.4 \\
 & & $\mu^{+}$ & 55.9 & 50.0 \\
 & & $\mu^{-}$ & 44.1 & 50.0 \\
200 & & $e^{+}$ & 63.9 & 49.7 \\
 & & $e^{-}$ & 35.6 & 48.3  \\
 & & $\mu^{+}$ & 63.7 & 50.1 \\
 & & $\mu^{-}$ & 35.7 & 48.1 \\ \hline
\end{tabular}
\end{center}
\caption{Relative number of entries, as percentages,  of the lepton
energy distributions \enplu\ and \enmin\ for $10^4$ event
simulations. For the 40 TeV 
simulations, case (a) refers to $n_B\sim 1/\alpha_W$, case
(b) \LOME\ prescription for $n_B$.}\label{tab:relheight}      
\end{table}

\begin{table}
\begin{center}
\begin{tabular}{|rl|r|r|} \hline
\multicolumn{4}{|c|}{No. events rel. occurrence} \\ \hline \hline
\multicolumn{2}{|c|}{Energy/TeV}& $\mu^{\pm}$ & $e^{\pm}$ \\ \hline
17 & & 340 & 285 \\
40 & (a) & 150 & 145\\
40 & (b) & 1130 & 1375 \\
200 & & 170 & 180 \\ \hline
\end{tabular}
\end{center}
\caption{Number of events for 95\% confidence level
by measuring relative occurrence of $l^{+}$ and $l^{-}$ as the
highest energy lepton of an event, compared
with the multi--W background predictions.}
\label{tab:lepoccur}      
\end{table}

The \BLNV\ leptons distort the shape of the \enplu\ distributions
compared to the \enmin\ ones. Thus, another experimental
test is the comparison of these spectra.
Within the context of our model, we may
estimate how many events are needed to see an asymmetry in the average
values of these spectra, by mixing \BLNV\ and multi--W data in
the same way as before. The estimates for the number of events are
given in Table~\ref{tab:lepensum}. It should 
be stressed again, however, that
the energy spectra are model dependent and so any attempt to verify
an asymmetry based on analysing the lepton spectra would be  
less reliable than simple lepton counting.

\begin{table}
\begin{center}
\begin{tabular}{|rl|r|r|} \hline
\multicolumn{4}{|c|}{Energy asymmetry} \\ \hline \hline
\multicolumn{2}{|c|}{Energy/TeV}& 
\multicolumn{1}{|c|}{Electron} & 
\multicolumn{1}{|c|}{Muon} \\ \hline
17 & & $3.7\ 10^3$ & $1.7\ 10^4$\\
40 & (a) & $6.0\ 10^2$ & $9.5\ 10^2$ \\
40 & (b) & $3.5\ 10^4$ & $2.2\ 10^4$ \\
200 & & $7.6\ 10^2$ & $7.1\ 10^2$ \\ \hline
\end{tabular}
\end{center}
\caption{Number of events for 95\% confidence level
by measuring asymmetry in energy
of highest energy lepton by event.}
\label{tab:lepensum}      
\end{table}

\begin{figure}[tb]
  \vspace*{5.9cm}
  \includegraphics{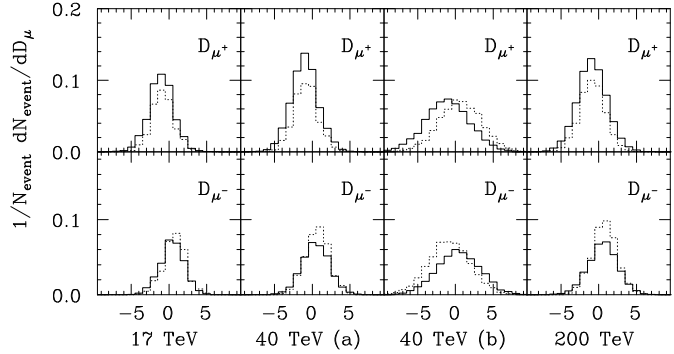}
  \caption[]{Difference in muon number.}
\label{fig:muon_diff}                     
\end{figure}

\begin{figure}[tb]
  \vspace*{5.9cm}
  \includegraphics{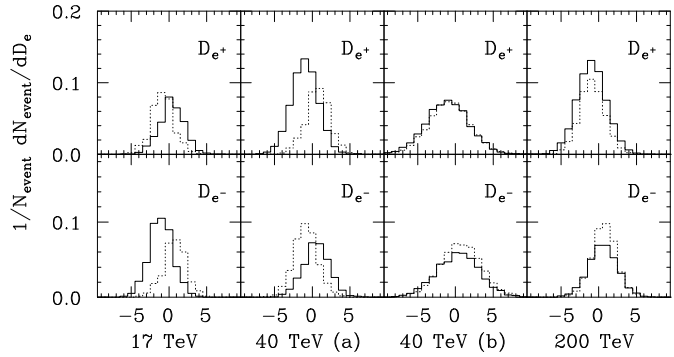}
  \caption[]{Difference in electron number.}
\label{fig:elec_diff}
\end{figure}

With this reservation in mind, we now consider refining the lepton
counting technique. We define two distributions, $D_{l}^{+}$ 
and $D_{l}^{-}$, by separating the measurement of $D_l$ according
to the sign of the highest energy lepton ($l^+$ or
$l^-$) on an event-by-event basis.
These distributions are plotted in Fig.~\ref{fig:muon_diff} for 
the muons and Fig.~\ref{fig:elec_diff} for
the electrons. The corresponding average values of $D_{l}^{+}$ and $D_{l}^{-}$ 
for the multi--W case are also non-zero, due mainly to the bias introduced by
using the lepton charge sign for event selection. The averages of these
distributions can be used as a signature of L violation, and estimates
of the number of events required are 
shown in Table~\ref{tab:lepenhan}.
Note that these estimates refer to
the {\em total} number of events. Despite the enhancement of
the signal due to the cut on the lepton charge, the total number
of events required is comparable to working with $D_l$ alone, due to
the fact that we have a reduced sample size.
\begin{table}
\begin{center}
\begin{tabular}{|rc|r|r|r|r|} \hline
\multicolumn{6}{|c|}{Number difference events} \\ \hline \hline
\multicolumn{2}{|c|}{Energy/TeV}& 
\multicolumn{1}{|c|}{$D_{e}^{+}$} & 
\multicolumn{1}{|c|}{$D_{e}^{-}$} & 
\multicolumn{1}{|c|}{$D_{\mu}^{+}$} &
\multicolumn{1}{|c|}{$D_{\mu}^{-}$} \\ \hline 
17 & & $1.1~10^4$ & $7.5~10^3$ & $9.4~10^3$ & $1.1~10^4$ \\
40 & (a) & $2.0~10^4$ & $2.4~10^4$ & $9.8~10^3$ & $2.9~10^4$ \\
40 & (b) & $1.4~10^4$ & $1.1~10^4$ & $1.7~10^4$ & $2.0~10^4$ \\
200 & & $1.6~10^4$ & $2.5~10^4$ & $4.7~10^4$ & $4.0~10^4$ \\ \hline
\end{tabular}
\end{center}
\caption{Number of events for demonstrating \BLNV\ at a 95\%
confidence level by measuring lepton number differences with
samples enhanced by the sign of the highest energy lepton.}
\label{tab:lepenhan}      
\end{table}


Some comments on the experimental considerations of detecting
leptons are in order. The large number of particles present
in the detector as a result of a \BLNV\ process will
make the identification of the particles difficult.
As discussed in Sect.~\ref{sec:blback}, there is little chance of
identifying individual jets or reconstructing the bosons from
the underlying hard process. However, it may be possible to
reconstruct leptonic $Z^0$ decays. This would reduce the number of
leptons in the $D_l$ analyses, and therefore should decrease the
size of the fluctuations. In addition, the $Z^0$ momentum 
spectrum could then be obtained from the reconstruction
process. However, a more detailed simulation of the measurement
of lepton energy and momentum is required to investigate this
further.

The detection criteria used are very
simplistic, and any attempt to detect \BLNV\ experimentally would
require a more complete detector simulation.
In particular, the isolation of electrons is complicated
by the presence of many nearby particles, which may `fake' an
electron in the detector. On the other hand, 
muons are normally identified
at the outer edge of a detector after having
passed through the intervening material. Therefore, muon
identification will probably be easier than electron 
identification.

It may be possible to alleviate some of the detection problems by
considering only isolated leptons.
Repeating the analysis based upon $D_{e}$ and $D_{\mu}$ for
the 40 TeV (\LOME\ prescription 
for {\nw}) simulation leads to an estimate 
of $\sim 10^4$ events for a 95\%
confidence level in verifying L violation. The larger number
of events required is due to the rejection of the non-isolated
leptons, which amount to approximately 80\%, 
reducing the statistics considerably.
The other three simulations, being
based on a $n_{B}\sim 1/\alpha_W$ approximation, have on average
less leptons per event; this is offset by the fact that there are
fewer particles in the detector. In order to investigate isolated
leptons further, a more detailed study, ideally including a 
full detector simulation, is required. 

\section{Conclusions}
\label{sec:conc}

We have discussed the construction, implementation and study of
a phenomenological model for \BLNV\ processes. The 
characteristic boson multiplicities and energies which are a 
feature of calculations based upon instanton and sphaleron
processes within the Standard Model have been incorporated.

The high boson multiplicities mean that such processes are very
distinctive from an experimental point of view. Particular
characteristics of the processes are their high threshold
energy, leading to `central' events in the detector, the spherical
nature of the interactions, and the
high particle multiplicities. The high transverse energy of the
final state, combined with the large number of high $p_T$ leptons and
the inability to resolve jets means that any conventional Standard
Model background can be easily separated. We have argued that the only
significant background is that of non-perturbative multi-W(Z) 
production, with typical boson multiplicities $\sim 1/\alpha_W$.

In our calculations we have taken into account the breaking of the electroweak
symmetry. We expect that the approximate
ratio of outgoing final state bosons in
the \BLNV\ process is $1 : 1 : {\cos}^2\theta_W : {\sin}^2\theta_W$ for
the $W^+$, $W^-$, $Z^0$ and $\gamma$ particles respectively. This has
allowed the study of the production of photons in \BLNV\ interactions
for the first time. The high energy and high $p_T$ regions of the
photon spectra are dominated by these photons, indicating that
the boson distributions from the \BLNV\ matrix elements 
can be directly studied. This study can be enhanced by including
reconstruction of leptonic $Z^0$ decays if possible.

The studies described here confirm that the demonstration of
baryon number violation is effectively impossible. 
The `extra' antibaryons from the B violating vertex are in
general produced at low energy and $p_T$, which makes them
difficult to isolate from the large number of other hadrons
produced. In addition, the charge of a baryon or antibaryon is
not directly related to the baryon number of the particle.
There is also the added problem of the low efficiency of baryon
identification at a supercollider. We quantify our pessimistic
conclusion by estimating the number of \BLNV\ (and multi-W background)
events required to verify B violation as at least $10^8$.

On the other hand, the studies show that the
verification of lepton number violation is possible, provided that
the cross section of the \BLNV\ process is high enough. Charged leptons
are relatively easy to identify at a supercollider, and there is the
advantage that the lepton number of a charged lepton is proportional
to its charge. A simple model independent test of L violation is to
count the number of $l^+$ and $l^-$ produced on an event-by-event
basis, and demonstrate an incompatability with L conservation. We find
that $\sim 10^3$ events would be required.  

We have also considered the enhancement of the sample of
\BLNV\ events compared to the background.
In particular, the momentum
distribution of the antileptons from the \BLNV\ process 
differs from that of
the leptonic decay products of the other outgoing particles.
Our studies show that using this
difference, in general, reduces the number of events one needs to
demonstrate an asymmetry, to $\sim 10^2$ events. It must be
stressed, however, that any enhancement of the sample using this
requires some knowledge of the momentum distribution of the
outgoing particles from the \BLNV\ interaction. Therefore, any
such study is not model independent.

The predictions made from these studies can be improved in a
number of ways. The most obvious is to improve the understanding
of the momentum distribution and angular structure of the outgoing state. 
Calculation of these quantities for the 
instanton matrix element (or any similar non-perturbative process
with a high boson multiplicity) is at present not possible due
to the large number of terms. Some insight may be gained by the type of
approach used to study sphaleron decays, which we saw are similar to
our model predictions in the corresponding region of parameter space.

We close with some comments on the future prospects for
studying not only \BLNV\ but non-perturbative processes in general.
One possibility is the observation of multi-W processes in
cosmic ray interactions~\cite{dm91,dm93a}. In 
particular, constraints can be placed
on the cross section for such processes in advance of any 
supercollider study. Secondly, there is theoretical 
evidence~\cite{kh91,ba93} that
instanton processes in \QCD\ may be observed in deep inelastic
$ep$ scattering at {\HERA}~\cite{ba93a}. 

\section*{Appendix A - The HERBVI event generator}

The \HERBVI\ event generator~\cite{gi94a} is 
a package written as an extension
to the \MC\ program \HW, and is based on the
\MAMBO\ \cite{mambopre} event generation algorithm.
Versions of \HW\ from  5.7 onwards contain hooks to allow the use
of this package.
The interface to \HW\
consists of four routines:
\begin{description}
\item[HVINIT] Initialization routine. This sets
up the physical parameters for the simulation. Once it has been
called, the user can modify these parameters. If it has not 
been called
before the first call to {\tt{HVHBVI}}, it is executed then.
\item[HVCBVI] Clustering routine. This is executed
within \HW\ to enable the antiquarks produced in the baryon number
violating process to be joined up according to  the colour constraints
described in Sect.~\ref{sec:colint}.
\item[HVHBVI] {Main event generator, called by \HW\ for
process codes {\tt IPROC} 7000 to 7999. At present, two processes are
generated by this routine, \\
\begin{tabular}{cl}
7000 & \BLNV\ event generation \\
7100 & Multi--W event generation \\
\end{tabular} }
\item[HVANAL] Initial event analysis. This routine produces a table
of all final state particles and a flag to indicate if the particle is
from the initial \BLNV\ interaction. The rapidity and
transverse momentum of each final state particle is calculated.
\end{description}

In addition, provision has been made for user generation of
both the parton-level cross section and the boson number distribution.
The routines {\tt{HUSGEN}} (parton-level cross section)
and {\tt{HURBOS}} (boson number distribution)
are called if the appropriate flags ({\tt HUFSFG} and 
{\tt HUFRBN} respectively) are set {\tt TRUE}.

Final-state particle momenta are generated using 
the \MAMBO\ algorithm, and to each event a weight $w$ is assigned.
The \HW\ package
produces an unweighted distribution by
accepting events for which the ratio $w/w_{max}$
exceeds a random number between $0$ and $1$. The maximum
weight $w_{max}$ is determined 
by a search at the start of the simulation.

In order to generate events with a high
efficiency, two simplifications are made.
Firstly, for each event only a {\em relative} weight is calculated.
This removes the need to calculate the phase space integral
for a given configuration. This integral is constant for a 
fixed boson multiplicity, and the 
relative weight for different \nw\ is taken into account when
the boson multiplicity distribution is generated.
Secondly, events are generated isotropically, as discussed above.
These two steps reduce the `tails' of the weight distribution,
and consequently increase the efficiency of the {\MC}.
It is straightforward to include the effect of a non-trivial
matrix element by reweighting the events generated by
{\HERBVI}. However, this will decrease the efficiency.

Use of the \HERBVI\ package requires that the dummy routines
{\tt{HVCBVI}} and {\tt{HVHBVI}} be deleted from the \HW\ package,
and that both programs are compiled with larger common blocks.
We find that a 10000 event simulation
takes approximately 5 to 10 hours to run on a Vax4000 
workstation. About a third of this time is used by \HERBVI\ when
generating events. More details of 
the \HERBVI\ package are available from
the authors, who can be contacted by electronic mail at
{\tt{gibbs@cbhep.cern.ch}} and {\tt{webber@vxcern.cern.ch}}.

\section*{Appendix B - The Student's t test}

The Student's $t$\ test is a standard tool in statistical analysis
for estimating the significance of a particular 
set of measurements, or that of
the difference between two sets of measurements.
Let $\bar{x}_1$ and $\bar{x}_{2}$, with variances $\sigma_1^2$ and
$\sigma_2^2$, be the result of two 
sets of measurements with $n_1$ and $n_2$
points in each set respectively.
We define the combined variance as
\begin{equation}
\sigma_{t}^{2} = \left( \frac{\sigma_{1}^{2}}{n_{2}}
+ \frac{\sigma_{2}^{2}}{n_{1}} \right) 
\frac{n_{1}+n_{2}}{n_{1}+n_{2}-2}
\end{equation}
with the number of degrees of freedom
\begin{equation}
N = n_{1} + n_{2} - 2.
\end{equation}
The estimate of the significance is then performed by
using the Student's $t$ distribution by evaluating
\begin{equation}
t = \frac{{\bar{x_{1}}} - {\bar{x_{2}}}}{\sigma_{t}}
\end{equation}
and computing the confidence level
\begin{equation}
F\left( t,N\right) = \frac{\Gamma\left(\frac{1}{2}\left( N+1\right)
\right)}{\sqrt{\pi N}\Gamma\left(\frac{1}{2}\right) }
\int_{-\infty}^{t} \left( 1 + \frac{x^2}{N}
\right)^{-\frac{1}{2}\left( N+1\right)}
\mathrm{d}x .
\end{equation}
F is then an estimate of the statistical significance of the difference
between the two sets of measurements. A fuller discussion of
this test is given in \cite{nrec}.


\begin{thebibliography}{10}

\bibitem{ri90}
{ A. Ringwald},
\newblock Nucl. Phys. B330 (1990) 1.

\bibitem{es90}
{ O. Espinosa},
\newblock Nucl. Phys. B343 (1990) 310.

\bibitem{mc90}
{L.~McLerran, A.~Vainshtein and M.~Voloshin},
\newblock Phys. Rev. D42 (1990) 171.

\bibitem{co90}
J.~Cornwall,
\newblock Phys. Lett. B243 (1990) 271.

\bibitem{gp90}
H.~Goldberg,
\newblock Phys. Lett. B246 (1990) 445.

\bibitem{gp92}
H.~Goldberg,
\newblock Phys. Rev. D45 (1992) 2945.

\bibitem{ma92}
M.~Mattis,
\newblock Phys. Rev. 214 (1992) 159.

\bibitem{tin92}
{P.~Tinyakov},
\newblock Int.J.Mod.Phys. A8 (1993) 1823.

\bibitem{ri93}
A.~Ringwald,
\newblock CERN preprint CERN-TH.6862 (1993).

\bibitem{ri91b}
{A.~Ringwald, F.~Schrempp and C.~Wetterich},
\newblock Nucl. Phys. B365 (1991) 3.

\bibitem{fa90}
{ G. Farrar and R. Meng},
\newblock Phys. Rev. Lett. 65 (1990) 3377.

\bibitem{gi94a}
{M.J.~Gibbs and B.R.~Webber},
\newblock in preparation.

\bibitem{ad69}
{S.\ Adler},
\newblock Phys. Rev. 177 (1969) 2426.

\bibitem{bj69}
{J.\ Bell and R.\ Jackiw},
\newblock Nuovo Cimento 51 (1969) 47.

\bibitem{ba69}
{W.A.\ Bardeen},
\newblock Phys. Rev. 185 (1969) 1848.

\bibitem{th76a}
{ G. 't Hooft},
\newblock Phys. Rev. D14 (1976) 3432.

\bibitem{th76b}
{G. 't Hooft},
\newblock Phys. Rev. Lett. 37 (1976) 8.

\bibitem{jr76}
{R.\ Jackiw and C.\ Rebbi},
\newblock Phys. Rev. Lett. 37 (1976) 172.

\bibitem{cdg76}
{C.\ Callan, R.\ Dashen and D.\ Gross},
\newblock Phys. Lett. B63 (1976) 334.

\bibitem{ma83}
N.~Manton,
\newblock Phys. Rev. D28 (1983) 2019.

\bibitem{km84}
F.~Klinkhammer and N.~Manton,
\newblock Phys. Rev. D30 (1984) 2212.

\bibitem{ya90}
L.~Yaffe,
\newblock in: Proc. of the Santa Fe Workshop on Baryon Number Violation at the
  SSC?, eds. M.~Mattis and E.~Mottola. (World Scientific, Singapore, 1990).

\bibitem{ar90}
P.~Arnold and M.~Mattis,
\newblock Phys. Rev. D42 (1990) 1738.

\bibitem{krt91}
{S.~Khlebnikov, V.~Rubakov and P.~Tinyakov},
\newblock Nucl. Phys. B350 (1991) 441.

\bibitem{mu91a}
A.H. Mueller,
\newblock Nucl. Phys. B348 (1991) 310.

\bibitem{mu91b}
A.H. Mueller,
\newblock Nucl. Phys. B353 (1991) 44.

\bibitem{vo91}
M.~Voloshin,
\newblock Nucl. Phys. B359 (1991) 301.

\bibitem{za90}
V.~Zakharov,
\newblock Minnesota preprint TPI-MINN-90/7-T (1990).

\bibitem{po90}
M.~Porrati,
\newblock Nucl. Phys. B347 (1990) 371.

\bibitem{kr91}
V.V. Khoze and A.~Ringwald,
\newblock Nucl. Phys. B355 (1991) 351.

\bibitem{dpe91}
D.~Diakonov and V.~Petrov,
\newblock in: Proc. of the 26th Winter School of the Leningrad Nuclear Physics
  Institute (1991).

\bibitem{mu91c}
A.H. Mueller,
\newblock Nucl. Phys. B364 (1991) 109.

\bibitem{ar91}
P.~Arnold and M.~Mattis,
\newblock Mod. Phys. Lett. A6 (1991) 2059.

\bibitem{dpo91}
D.~Diakonov and M.~Polyakov,
\newblock St. Petersburg preprint LNPI-1737 (1991).

\bibitem{bs92}
{I.~Balitskii and A.~Sch\"afer},
\newblock Nucl. Phys. B404 (1993) 639.

\bibitem{sil92}
{P.G.~Silvestrov},
\newblock Phys. Lett. B323 (1994) 25.

\bibitem{za92a}
V.~Zakharov,
\newblock Nucl. Phys. B371 (1992) 637.

\bibitem{za91}
{V.I.\ Zakharov},
\newblock Nucl. Phys. B353 (1991) 683.

\bibitem{ms91}
{M.\ Maggiore and M.\ Shifman},
\newblock Nucl. Phys. B365 (1991) 161.

\bibitem{ve92}
{G.~Veneziano},
\newblock Mod. Phys. Lett. A7 (1992) 1661.

\bibitem{ag87}
{H.\ Aoyama and H.\ Goldberg},
\newblock Phys. Lett. B188 (1987) 506.

\bibitem{amc88}
{P.\ Arnold and L.\ McLerran},
\newblock Phys. Rev. D37 (1988) 1020.

\bibitem{he92}
M.~Hellmund and J.~Kripfganz,
\newblock Nucl. Phys. { B373} (1992) 749.

\bibitem{za92}
J.~Zadrozny,
\newblock Phys. Lett. B284 (1992) 88.

\bibitem{ra88}
B.~Ratra and L.G. Yaffe,
\newblock Phys. Lett. {B205} (1988) 258.

\bibitem{wi77}
E.~Witten,
\newblock Phys. Rev. Lett. {38} (1977) 121.

\bibitem{tu91}
N.~Turok and J.~Zadrozny, 
\newblock Nucl. Phys. {B358} (1991) 471.

\bibitem{ak89}
{T.~Akiba, H.~Kikuchi, and T.~Yanagida},
\newblock Phys. Rev. {D40} (1989) 588.

\bibitem{kl68}
J.R. Klauder and E.C.G. Sundarshan,
\newblock {\em {Fundamentals of Quantum Optics}}.
\newblock Benjamin, New York, 1968.

\bibitem{zh90}
{W.-M.~Zhang, D.H.~Feng, and R.~Gilmore},
\newblock Rev. Mod. Phys. {\bf 62} (1990) 867.

\bibitem{uzi}
C.~Itzykson and J.~Zuber,
\newblock {\em Quantum Field Theory}.
\newblock McGraw Hill, 1985.

\bibitem{ehkq86}
{E.~Eichten, I.~Hinchliffe, K.~Lane and C.~Quigg},
\newblock Rev. Mod. Phys. 58 (1986) 1065.

\bibitem{Bykling}
E.~Bykling and K.~Kajantie,
\newblock {\em { Particle Kinematics }},
\newblock Wiley-Interscience Pub., 1971.

\bibitem{mambopre}
{R.~Kleiss and J.~Stirling},
\newblock Nucl. Phys. B385 (1992), 413.

\bibitem{rambo}
{W. J. Stirling, R. Kleiss and S. D. Ellis},
\newblock Comp. Phys. Comm. 40 (1986) 359.

\bibitem{gi94b}
{M.J.~Gibbs, A.~Ringwald and B.R.~Webber},
\newblock in preparation.

\bibitem{hw92}
{G.\ Marchesini, B.R.\ Webber, G.\ Abbiendi, I.G.\ Knowles, M.H.\ Seymour and
  L.\ Stanco},
\newblock Comp. Phys. Comm. 67 (1992) 465.

\bibitem{herwigphys}
{G. Marchesini and B.R. Webber },
\newblock Nucl. Phys. B310 (1988) 461.

\bibitem{js84}
{T.\ Sj\"{o}strand},
\newblock Comp. Phys. Comm. 39 (1986) 347.

\bibitem{js87}
{M.\ Bengtsson and T.\ Sj\"{o}strand},
\newblock Comp. Phys. Comm. 43 (1987) 367.

\bibitem{clus}
B.R.\ Webber,
\newblock Nucl. Phys. B238 (1984) 492.

\bibitem{string}
{B.~Andersson, G.~Gustafson, G.~Ingelman and T.~Sj\"ostrand},
\newblock Phys. Rep. 97 (1983) 33.

\bibitem{ang81a}
{A.H.\ Mueller},
\newblock Phys. Lett. 104 (1981) 161.

\bibitem{ang81b}
B.I.\ Ermolaev and V.S.\ Fadin,
\newblock JETP Lett. 33 (1981) 285.

\bibitem{QCDbook}
{Yu.L.\ Dokshitzer, V.A.\ Khoze, A.H.\ Mueller and S.I.\ Troyan},
\newblock {\em Basics of Perturbative QCD},
\newblock Editions Frontieres, 1991.

\bibitem{mw84}
G.\ Marchesini and B.R.\ Webber,
\newblock Nucl. Phys. B238 (1984) 1.

\bibitem{heavy}
G.\ Marchesini and B.R.\ Webber,
\newblock Nucl. Phys. B330 (1990) 261.

\bibitem{ba92}
{V.\ Barger, A.L.\ Stange and R.J.N.\ Phillips},
\newblock Phys. Rev. D45 (1992) 1484.

\bibitem{dm93}
D.A.\ Morris,
\newblock Talk at {\em Workshop on Physics at Current Accelerators and the
  Supercollider}, UCLA preprint 93/TEP/30.

\bibitem{gi92}
{W.T.\ Giele, T.\ Matsuura, M.H.\ Seymour and B.R.\ Webber},
\newblock in {\em Research Directions for the Decade}, Proc.\ 1990 Snowmass
  Summer Study on High Energy Physics, ed.\ E.L.\ Berger (World Scientific,
  Singapore, 1992).

\bibitem{be91}
{F.A.\ Berends, H.\ Kuijf, B.\ Tausk and W.T.\ Giele},
\newblock Nucl. Phys. B357 (1991) 32.

\bibitem{ch91}
{R.S.\ Chivukula, M.\ Golden and E.H.\ Simmons},
\newblock Nucl. Phys. B363 (1991) 83.

\bibitem{od90}
R.\ Odorico,
\newblock {Int.\ J.\ Mod.\ Phys.} A5 (1990) 3617.

\bibitem{go93}
H.\ Goldberg and R.\ Rosenfeld,
\newblock Northeastern University preprint NUB-3061/93-Th. (1993).

\bibitem{Atlas}
ATLAS collaboration,
\newblock {\em Letter of Intent},
\newblock CERN/LHCC/92-4.

\bibitem{CMS}
CMS collaboration,
\newblock {\em Letter of Intent},
\newblock CERN/LHCC/92-3.

\bibitem{nrec}
{W.H.~Press, S.A.~Teukolsky, W.T.~Vetterling and B.P.~Flannery},
\newblock {\em {Numerical Recipes in FORTRAN, Second Edition}},
\newblock Cambridge Univ. Press, 1992.

\bibitem{dm91}
{D.A.~Morris and R.~Rosenfeld}, 
\newblock Phys. Rev. D44 (1991) 3530.

\bibitem{dm93a}
{D.A.~Morris and A.~Ringwald},
\newblock Astroparticle Physics 2 (1994) 43.

\bibitem{kh91}
{V.V.~Khoze and A.~Ringwald},
\newblock Phys. Let. B259 (1991) 106.

\bibitem{ba93}
{I.I.~Balitsky and V.M.~Braun},
\newblock Phys. Lett. B314 (1993) 237.

\bibitem{ba93a}
{I.I.~Balitsky and V.M.~Braun},
\newblock Phys. Rev. D47 (1993) 1879.

\end{thebibliography}

\end{document}